\documentclass[10pt, twocolumn, letterpaper]{article}

\usepackage[pagenumbers]{iccv} % To force page numbers, e.g. for an arXiv version

\definecolor{iccvblue}{rgb}{0.21,0.49,0.74}
\usepackage[pagebackref, breaklinks, colorlinks, allcolors=iccvblue]{hyperref}
\usepackage{lineno}
\usepackage{multirow}
\usepackage{array}
\newcommand{\PreserveBackslash}[1]{\let\temp=\\#1\let\\=\temp}
\newcolumntype{C}[1]{>{\PreserveBackslash\centering}p{#1}}
\newcolumntype{R}[1]{>{\PreserveBackslash\raggedleft}p{#1}}
\newcolumntype{L}[1]{>{\PreserveBackslash\raggedright}p{#1}}
\usepackage{graphicx}
\usepackage{color}
\usepackage{colortbl}
\usepackage{amsmath, amsfonts}
\usepackage{graphicx}
\usepackage{multirow}
\usepackage{mathrsfs}
\usepackage{mathtools}
\usepackage{amsthm}
\usepackage{multirow}
\usepackage{booktabs}
\usepackage{threeparttable}
\usepackage[misc]{ifsym}
\usepackage{enumerate}
\usepackage{enumitem}
\usepackage{bbding}
\usepackage{bbm}

\title{PIGUIQA: A Physical Imaging Guided Perceptual Framework for \\ Underwater Image Quality Assessment}

\author{Weizhi Xian $^{1,2}$
\and
Mingliang Zhou $^{3,*}$
\and
Leong Hou U $^{4}$
\and
Zhengguo Li $^{5}$
\and
\\
$^{1}$Chongqing Research Institute of Harbin Institute of Technology,\\Harbin Institute of Technology, Chongqing 401151, China\\
$^{2}$Faculty of Computing, Harbin Institute of Technology, Harbin, Heilongjiang 150001, China\\
$^{3}$School of Computer Science, Chongqing University, Chongqing 40044, China\\
$^{4}$Faculty of Science and Technology, University of Macau, Macau 519000, China\\
$^{5}$Institute for Infocomm Research, Agency for Science, Technology and Research, Singapore 138632\\
}

\begin{document}
\maketitle

\begin{abstract}
In this paper, we propose a Physical Imaging Guided perceptual framework for Underwater Image Quality Assessment (UIQA), termed PIGUIQA. First, we formulate UIQA as a comprehensive problem that considers the combined effects of direct transmission attenuation and backward scattering on image perception. By leveraging underwater radiative transfer theory, we systematically integrate physics-based imaging estimations to establish quantitative metrics for these distortions. Second, recognizing spatial variations in image content significance and human perceptual sensitivity to distortions, we design a module built upon a neighborhood attention mechanism for local perception of images. This module effectively captures subtle features in images, thereby enhancing the adaptive perception of distortions on the basis of local information. Third, by employing a global perceptual aggregator that further integrates holistic image scene with underwater distortion information, the proposed model accurately predicts image quality scores. Extensive experiments across multiple benchmarks demonstrate that PIGUIQA achieves state-of-the-art performance while maintaining robust cross-dataset generalizability. The implementation is publicly available at \href{https://github.com/WeizhiXian/PIGUIQA}{https://github.com/WeizhiXian/PIGUIQA}.
\end{abstract}

\section{Introduction}
\label{sec:intro}

The underwater world is rich in resources, and underwater images provide a crucial medium for exploring environments such as oceans and lakes by accurately and intuitively capturing underwater information \cite{46}. However, the complexities of the underwater imaging environment often result in suboptimal image quality \cite{45,18,19}. Therefore, underwater image quality assessment (UIQA) is fundamental for underwater image processing. UIQA methods help determine whether underwater images meet quality requirements for various applications and assess the visual quality of enhanced underwater images.

According to the accessibility of ideal references, image quality assessment (IQA) methods can be broadly categorized into full-reference (FR) \cite{48}, reduced-reference (RR) \cite{49} and no-reference (NR) \cite{47} approaches. For underwater images, obtaining perfect undistorted references is often impractical; thus, typical FR-IQA approaches, such as the structural similarity index measure (SSIM) \cite{20} and feature similarity indexing method (FSIM) \cite{21}, are not applicable. Although NR-IQA approaches \cite{4,5,9} have been developed for many years, most of them are designed to evaluate generic images. Many NR-IQA methods \cite{10,26,24} are based on statistical properties, such as natural scene statistics (NSS), which are typically observed under good lighting conditions and in clear visual environments. However, the unique characteristics of underwater environments result in light propagation and scattering rules that are significantly different from those in terrestrial natural scenes. As a result, the statistical properties of NSS often fall short in accurately describing and assessing the quality of underwater images.

To date, effective, robust, and widely accepted UIQA methods in the field of underwater image processing are lacking. Although several well-known UIQA methods have been developed, such as underwater color image quality evaluation (UCIQE) \cite{34} and the underwater image quality measure (UIQM) \cite{35}, most UIQA methods are based on manual-crafted features, which often fail to fully capture the complexity and diversity of underwater imagery. In addition, these methods often prioritize images with higher color saturation, resulting in quality assessments that may not align with human visual perception. In addition to manually designed features, recent advancements in deep learning have demonstrated significant feature learning capabilities. However, only a few studies \cite{30} have applied deep learning to UIQA. This is because deep learning models typically require large amounts of labelled data for training, but the existing annotated underwater image datasets are relatively small. The data dependency in deep learning also makes ensuring the generalization capability of deep learning models challenging.

\begin{figure}[t]
  \centering
  \includegraphics[width=0.48\textwidth]{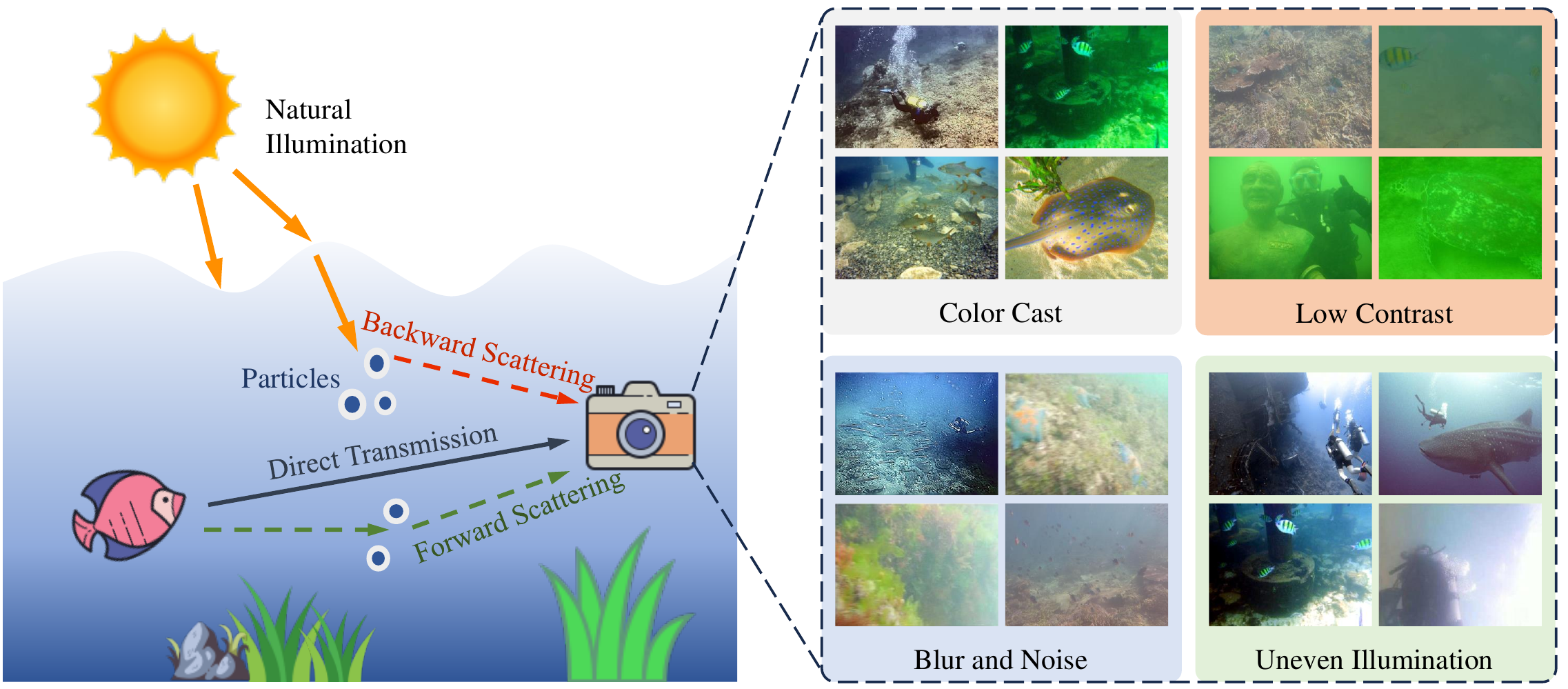}\\
  \caption{Illustration of the underwater optical imaging model with resulting major distortions. The scattering effects of light in water, caused by visible suspended particles and dissolved organic matter, result in varying degrees of deviation in the light transmission path. This alteration in the energy distribution of the light beam significantly degrades the quality of underwater images. The distorted images, which are sampled from various datasets, typically exhibit multiple types of distortions.}
  \label{fig:imaging}
\end{figure}

Another primary challenge is that underwater images differ markedly from terrestrial images because of their specific imaging environments and lighting conditions. As a result, traditional NR-IQA methods that perform well in terrestrial image processing cannot be directly applied to underwater images. Specifically, as shown in \textbf{Fig. \ref{fig:imaging}}, the underwater environment is complex and variable, with light propagation affected by scattering and absorption. Different wavelengths of light attenuate at different rates, with longer wavelengths diminishing more rapidly in water. Consequently, shorter wavelengths such as green and blue light can travel further underwater, causing underwater images to exhibit a blue–green hue and suffering from significant color distortion. The scattering of light in water can be categorized into forward scattering and backward scattering on the basis of the angle. Forward scattering occurs when light reflected from the target object deviates from its original path before it reaches the sensor, resulting in blurred underwater images. Backward scattering, on the other hand, involves a large amount of stray light entering the sensor, leading to visual issues such as loss of detail, low overall contrast, and increased background noise in underwater images. The need for the development of specialized UIQA methods that can effectively address the unique challenges posed by underwater imaging is urgent.

To address the aforementioned issues, combining prior knowledge of underwater imaging with the excellent feature extraction capabilities of deep learning presents an effective solution. On this basis, we propose a physically imaging guided framework for UIQA (PIGUIQA) by integrating the physical model of underwater light propagation with deep learning techniques. Specifically, the framework considers distortions caused by backward scattering and direct transmission processes. These factors play crucial roles in underwater imaging, influencing image clarity, contrast, and color representation. Within the PIGUIQA framework, we leverage the powerful feature extraction capabilities of deep learning models to capture both local details and global perceptual features in images. Capturing local details allows the model to perceive distortions in critical regions of underwater images, such as object edges and texture information, whereas global perceptual features help the model understand the overall structure and context of the entire image. The main contributions of this work can be summarized as follows:
\begin{itemize}
\item We formulate UIQA as a comprehensive problem that accounts for the combined effects of direct transmission attenuation and backward scattering on image perception. We integrate advanced physics-based underwater imaging estimations into our framework. From the perspective of matrix theory, we define distortion metrics to measure these impacts.
\item We design a neighborhood attention (NA)-based local perceptual module to capture subtle features in images, thereby enhancing the adaptive perception of distortions via local information. Additionally, a global perceptual aggregator is employed to integrate the original image content with underwater image distortion information, assisting the model in understanding the overall structure and context of the entire image.
\item The experimental results demonstrate that the proposed method effectively evaluates underwater image quality, achieving superior performance across various correlation coefficients and error metrics. Furthermore, a cross-dataset experiment also confirms the strong generalizability and robustness of the proposed method.
\end{itemize}

% The remainder of this paper is organized as follows. In \textbf{Section \ref{sec:related}}, we present some related work. The proposed method is described in detail in \textbf{Section \ref{sec:method}}. In \textbf{Section \ref{sec:exp}}, we conduct comprehensive experiments. Finally, the conclusions of this paper are given in \textbf{Section \ref{sec:conclu}}.

\section{Related Work}
\label{sec:related}

Owing to the unique characteristics of underwater images, conventional NR-IQA methods \cite{4,6,7,38} are not suitable for evaluating underwater images. Consequently, researchers have begun to focus on specially designed UIQA methods, and many important studies have emerged in this field. Yang \etal \cite{34} proposed the underwater color image quality evaluation method (UCIQE), which quantifies quality attributes such as color bias, blurriness, and contrast through a linear combination of chromaticity, saturation, and contrast measurements. Panetta \etal \cite{35} introduced an underwater image quality measure (UIQM) from the perspective of the human visual system (HVS), which considers colorfulness, sharpness, and contrast, thereby addressing some limitations of the UCIQE. Guo \etal \cite{44} also focused on these three features and conducted experiments on a self-constructed small-scale dataset containing only 200 underwater enhanced images, which may restrict its performance on larger datasets. Wang \etal \cite{17} developed a UIQA method called the CCF that incorporates factors such as colorfulness, contrast, and fog density. Yang \etal \cite{14} introduced a frequency domain UIQA metric (FDUM), which, like UIQM, considers colorfulness, contrast, and sharpness but analyses these features in the frequency domain and integrates the dark channel prior (DCP) \cite{39}. Jiang \etal \cite{29} proposed a no-reference underwater image quality (NUIQ) metric that transforms underwater images from the RGB space to the opponent color space (OC); extracts color, luminance, and structural features in the OC space; and employs support vector machines (SVMs) for quality assessment. Zheng \etal \cite{40} introduced the underwater image fidelity (UIF) metric, which evaluates the naturalness, sharpness, and structural indicators in the CIELab color space. Guo \etal \cite{41} presented an underwater image enhancement quality metric (UWEQM) that considers features such as transmission medium maps, Michaelson-like contrast, salient local binary patterns, and simplified color autocorrelograms. Liu \etal \cite{3} developed an underwater image quality index (UIQI) by extracting features related to luminance, color cast, sharpness, contrast, fog density, and noise, using support vector regression (SVR) to predict image quality. However, these methods, which rely on handcrafted features and regression techniques, still have limitations, as they do not comprehensively characterize image quality and may not be effective for various types of underwater images.

With the rapid advancement of deep neural networks, there is an urgent need for end-to-end UIQA methods that can automatically learn useful features from data and predict image quality more accurately. Fu \etal \cite{42} proposed a rank learning framework, which was the first to utilize deep learning for UIQA. This method generates a set of medium-quality images by blending original images with their corresponding reference images at varying degrees and employs a Siamese network to learn their quality rankings. Wang \etal \cite{43} introduced a generation-based joint luminance-chrominance underwater image quality evaluation (GLCQE) method. GLCQE first employs DenseUNet to generate two reference images—one unenhanced and one optimally enhanced—and uses these images to assess chromatic and luminance distortions while also designing a parallel spatial attention module to represent image sharpness. Recognizing that deep learning often requires large datasets for training, Jiang \etal \cite{30} constructed a dataset for UIQA with multidimensional quality annotations and proposed a multistream collaborative learning network (MCOLE). This method trains three specialized networks to extract color, visibility, and semantic features, facilitating quality prediction through multistream collaborative learning.

Despite the significant progress made by these deep learning methods in UIQA, they often lack consideration of the underwater imaging process and may only be effective on specific datasets, limiting their generalization capabilities. Therefore, integrating prior knowledge of the underwater imaging process with deep learning techniques is an effective strategy for addressing these challenges.

\section{Methodology}
\label{sec:method}

\subsection{Problem formulation}
\label{subsec:prob_formu}

\begin{figure*}[t]
  \centering
  \includegraphics[width=0.99\textwidth]{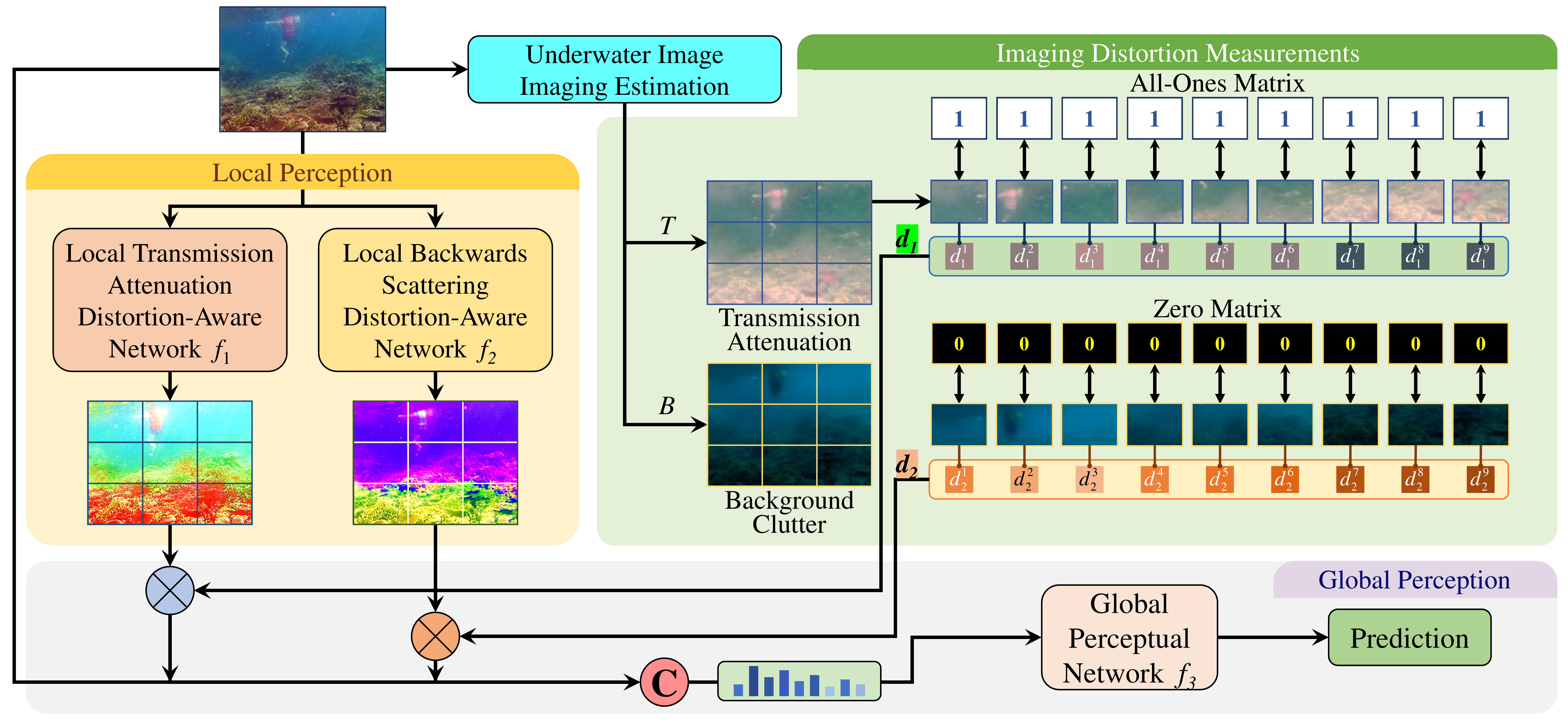}\\
\caption{Overall structure of the proposed UIQA method. The symbols \normalsize``\large{$\otimes $}\normalsize'' and ``\normalsize{\textcircled{\scriptsize{C}}}\normalsize'' denote the Hadamard product and concatenation operations, respectively. }
\vspace{-5pt}
\label{fig:UIQA}
\end{figure*}

The problem of IQA can be formulated as the task of finding a function $f$ that produces evaluation results closely aligned with the true quality of an image. This can be expressed mathematically as follows:
\begin{equation}\label{eq:iqa}
  \mathop {\min }\limits_f \mathbb{E}\left\{ Q - f(I) \right\}
\end{equation}
where $Q$ represents the ground truth quality of an image $I$, typically expressed in terms of the mean opinion score (MOS). $\mathbb{E}$ denotes the mathematical expectation. This problem involves identifying a function $f$ capable of accurately reflecting various image attributes, such as sharpness, contrast, color, and noise levels, to derive an accurate quality score.

For underwater images, we specifically analyse image quality by considering the causes of distortion inherent to underwater imaging. In real underwater scenarios, the camera is often positioned close to the underwater scene. Consequently, the impact of forward scattering on image quality can be neglected, resulting in only direct transmission attenuation and backward
scattering components to be considered. This relationship is represented as:
\begin{equation}\label{eq:imaging1}
 I \triangleq I_{dis} = I_{per} \otimes T + B
\end{equation}
where $I_{per}$ represents the perfect image, $I_{dis}$ represents the corresponding distorted image, $T$ denotes the transmission attenuation matrix, and $B$ denotes the background clutter matrix. The symbol $\otimes$ denotes the Hadamard product.

Both $T$ and $B$ can be estimated via traditional modelling methods that are based on prior knowledge \cite{36} or by employing deep learning models trained on synthetic underwater image datasets \cite{1}. The all-ones matrix is denoted as $E$, and the zero matrix is denoted as $O$. For a high-quality image, the direct transmission attenuation matrix $T$ should be ``closer'' to $E$, whereas the background noise matrix $B$ should be ``closer'' to $O$. Thus, we introduce the concept of imaging-perceived distortion, which is directly correlated with the distances between $T$ and $E$ and between $B$ and $O$. Given that perceived distortion is closely related to the image content itself, the UIQA problem can be formulated as follows:
\begin{equation}\label{eq:problem}
  \mathop {\min }\limits_{{f_1},{f_2},{f_3}} \hspace{-4pt}\mathbb{E}\hspace{-2pt} \left\{ {{{Q}} \hspace{-2pt}-\hspace{-2pt} f_3\hspace{-1pt}\big(I\hspace{-2pt}, \hspace{-1pt}{{f_1}\hspace{-1pt}({I}) \hspace{-2pt}\otimes\hspace{-2pt} {d_1}\hspace{-1pt}(T\hspace{-1pt},\hspace{-1pt}E)\hspace{-1pt},\hspace{-1pt}{f_2}\hspace{-1pt}({I}) \hspace{-2pt}\otimes\hspace{-2pt} {d_2}\hspace{-1pt}(B\hspace{-1pt},\hspace{-1pt}O)} \big)} \right\}
\end{equation}
where $Q$ is the ground-truth quality of an underwater image and where $d_1$ and $d_2$ represent the distance maps measuring transmission attenuation and backward scattering distortion between $T$ and $E$ and between $B$ and $O$, respectively. $f_1$ and $f_2$ are local distortion-aware functions to be learned, and $f_3$ is a global perceptual function to be learned. This formulation captures the essential aspects of UIQA by integrating a physical model into the evaluation process.

\subsection{Framework}
\label{subsec:frame}

\textbf{Fig. \ref{fig:UIQA}} illustrates the proposed framework for UIQA. The process begins with an underwater image, which is processed through an image estimation module that generates two key maps: the transmission attenuation map $T$ and the background clutter map $B$. Transmission attenuation and background clutter distortions are quantified by evaluating the distance between each patch of $T$ and an all-one matrix and the distance between each patch of $B$ and a zero matrix, respectively. These distortions are denoted as $d_1$ and $d_2$. The local perceptual modules $f_1$ and $f_2$ subsequently capture local scattering and transmission distortion characteristics. After local processing, the outputs of these modules are scaled by $d_1$ and $d_2$ and then fused with the original image. This combined data is then passed through a global perceptual network $f_3$, which integrates the information from all previous stages to generate a final prediction. The proposed framework effectively combines local distortion correction with global perceptual adjustment, which is tailored specifically for underwater optical scenarios.

\subsection{Underwater Imaging Distortion Measurements}
\label{subsec:dis}
 A precise and detailed version of Eq. (\ref{eq:imaging1}) is as follows \cite{2}:
\begin{equation}\label{eq:imaging2}
\begin{aligned}
  I_{dis}^c(x,y) &= I_{per}^c(x,y) \cdot {e^{ - \beta _D^c(x,y) \cdot l(x,y)}}\\
  &+ B_\infty ^c(x,y) \cdot \left({1 - {e^{ - \beta _B^c(x,y) \cdot l(x,y)}}} \right)
\end{aligned}
\end{equation}
where $c \in \{R,G,B\}$ is the color channel; $I_{per}$ and $I_{dis}$ are distortion-free and distorted images, respectively; $l$ is the object‒camera distance; $B_\infty$ is the background ambient light image; and $\beta _D$ and $\beta _B$ are the direct attenuation and backward scattering coefficients, respectively.

Based on Equation (2), many underwater image restoration methods \cite{36,1,45} utilizing imaging estimation have been proposed, which can be mathematically formulated as:
\begin{equation}\label{eq:syreanet}
\hat{T},\hat{B},\hat{I}_{per} =  \mathcal{F}(I_{dis})
\end{equation}
where $\mathcal{F}(\cdot)$ denotes the restoration model which can estimate the transmission attenuation matrix $T$ and background clutter matrix $B$, as defined in Eq. (\ref{eq:imaging1}). Given that the primary objective of this study is to propose an UIQA framework rather than developing underwater image restoration algorithms which constitutes a distinct and complex research topic in computer vision, we employ a pre-trained network to handle the imaging estimation component.

\begin{figure}[t]
\begin{minipage}{0.24\linewidth}
  \centering
\centerline{\includegraphics[width=0.095\paperwidth, height=0.085\paperwidth]{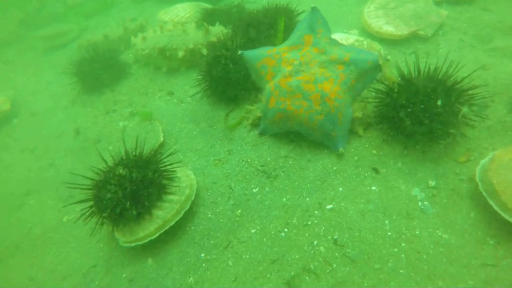}}
\vspace{2pt}
\end{minipage}
\hfill
\begin{minipage}{0.24\linewidth}
  \centering
\centerline{\includegraphics[width=0.095\paperwidth, height=0.085\paperwidth]{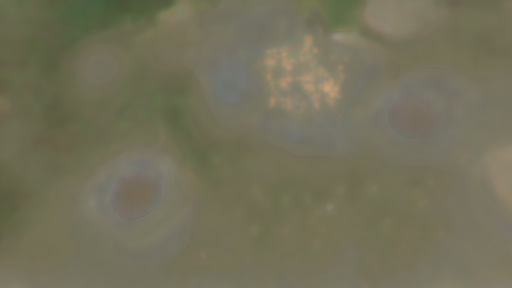}}
\vspace{2pt}
\end{minipage}
\hfill
\begin{minipage}{0.24\linewidth}
  \centering
\centerline{\includegraphics[width=0.095\paperwidth, height=0.085\paperwidth]{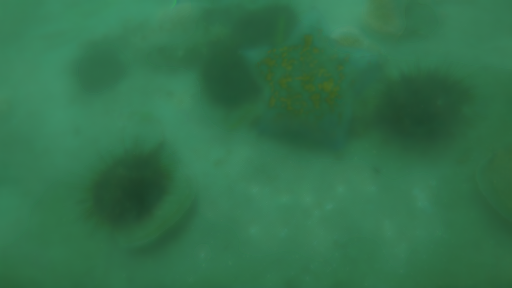}}
\vspace{2pt}
\end{minipage}
\hfill
\begin{minipage}{0.24\linewidth}
  \centering
\centerline{\includegraphics[width=0.095\paperwidth, height=0.085\paperwidth]{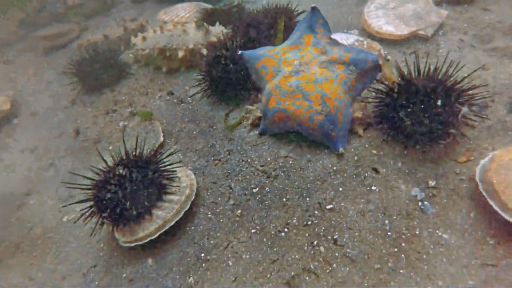}}
\vspace{2pt}
\end{minipage}

\begin{minipage}{0.24\linewidth}
  \centering
\centerline{\includegraphics[width=0.095\paperwidth, height=0.085\paperwidth]{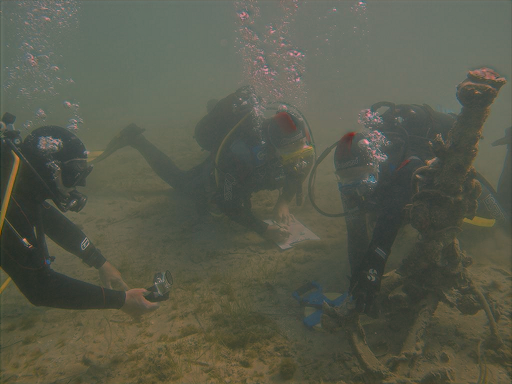}}
\centerline{\scriptsize{(a)}}
\end{minipage}
\hfill
\begin{minipage}{0.24\linewidth}
  \centering
\centerline{\includegraphics[width=0.095\paperwidth, height=0.085\paperwidth]{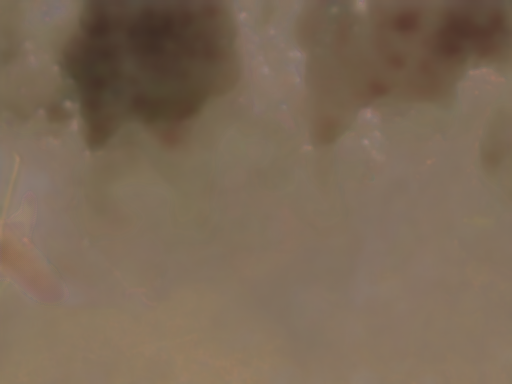}}
\centerline{\scriptsize{(b)}}
\end{minipage}
\hfill
\begin{minipage}{0.24\linewidth}
  \centering
\centerline{\includegraphics[width=0.095\paperwidth, height=0.085\paperwidth]{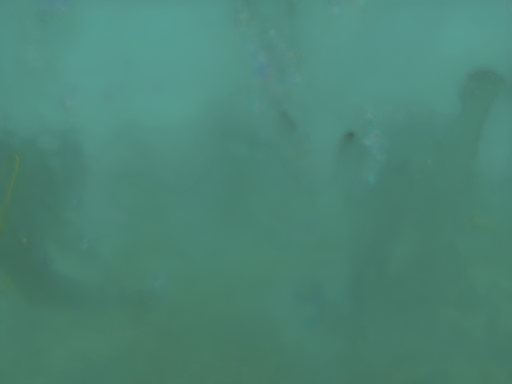}}
\centerline{\scriptsize{(c)}}
\end{minipage}
\hfill
\begin{minipage}{0.24\linewidth}
  \centering
\centerline{\includegraphics[width=0.095\paperwidth, height=0.085\paperwidth]{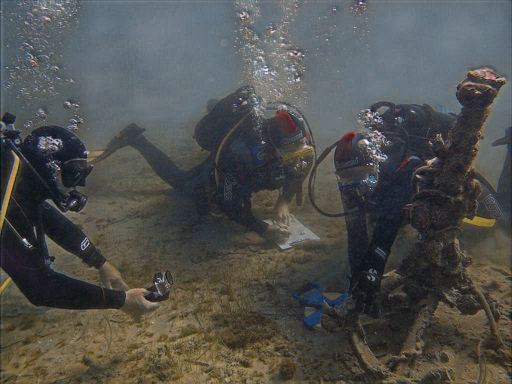}}
\centerline{\scriptsize{(d)}}
\end{minipage}

\caption{Illustration of underwater imaging estimation via Eq. (\ref{eq:syreanet}). (a): Input distorted underwater images $I_{dis}$. (b): Estimated transmission attenuation map $\hat{T}$. (c): Estimated background clutter map $\hat{B}$. (d): Restored underwater images $\hat{I}_{per}$. In the transmission attenuation map, darker areas signify more severe scattering and greater distortion, whereas in the background clutter map, brighter areas indicate a greater degree of background noise.}
\label{fig:img_est}
\end{figure}

As demonstrated in \textbf{Fig. \ref{fig:img_est}}, underwater imaging estimation enables quantitative characterization of distortion patterns, providing critical a priori information for subsequent image quality assessment. Considering the varying depths and optical properties of different regions and objects in underwater images, we decompose the image transmission attenuation map $T$ and the background clutter map $B$ into patches of size $N \times N$. This strategy facilitates the model's learning of local features, thereby enhancing the perception of details in images with intricate content. Consequently, we address the design of imaging distortion metrics for each patch.

For the transmission attenuation map $T$, which ideally should resemble the all-ones matrix $E$, we define a linear operator for any $N \times N$ matrix $A$ as follows:
\begin{equation}\label{eq:OT}
  \mathscr{T}^k(A) = (T^k - E_{N \times N}) \otimes A
\end{equation}
where $T^k$ is the $k$-th patch of $T$. On the basis of this definition, we establish the transmission attenuation distortion metric for a patch of size $N \times N$ as the norm of the linear transformation $\mathscr{T}$:
\begin{equation}\label{eq:d1_define}
  d_1^k = d_1(T^k, E_{N \times N}) = \|\mathscr{T}^k\|
\end{equation}
According to the functional analysis theory, we have:
\begin{equation}\label{eq:norm_T}
  \|\mathscr{T}^k\| = \sup_{\|A\|_2 \leq 1} \| \mathscr{T}^k(A) \|_2 = |\sigma_{\max}(\mathscr{T}^k)|
\end{equation}
where $\sigma_{\max}$ denotes the largest eigenvalue of the operator. Since
\begin{equation}\label{eq:eig}
  \mathscr{T}^k(E_{i,j}) = (T^k_{i,j} - 1)E_{i,j}
\end{equation}
where $T^k_{i,j}$ is the $(i,j)$-th element of $T^k$ and where $E_{i,j}$ is an $N \times N$ matrix with the $(i,j)$-th element as 1 and all other elements as 0, $T^k_{i,j} - 1$ serves as the eigenvalue of $\mathscr{T}^k$. Therefore, we define the transmission attenuation distortion measurement in the $k$-th patch as follows:
\begin{equation}\label{eq:d1}
  d_1^k = \mathop {\max }\limits_{1 \le i,j \le N} |T_{i,j}^k - 1|
\end{equation}

The background clutter map can be regarded as an additional noise overlay on the original image, arising from background textures, variations in lighting, or the presence of nontarget objects. According to the theory of human visual sensitivity, the HVS can only distinctly recognize distortion when the intensity of the noise surpasses a specific perceptual threshold. Thus, we propose a backward scattering distortion measurement for the $k$-th patch of $B$:
\begin{equation}\label{eq:d2}
  d_2^k= d_2 (B^k, O_{N \times N})= \mathop {\max }\limits_{1 \le i,j \le N} B_{i,j}^k
\end{equation}
where $B^k_{i,j}$ is the $(i,j)$-th element of $B^k$. The core idea of this measurement is that for each local patch, the degree of distortion can be quantified by the maximum noise value within that region, as the peak noise level typically generates the most significant visual interference, making it more readily detectable by the HVS. At this point, we can obtain $d_1$ and $d_2$ via Eq. (\ref{eq:problem}).

\subsection{Perceptual Networks}
\label{subsec:vie}

High-quality underwater images typically exhibit good contrast, smooth and consistent textures, and sharp edges in local regions. Therefore, analysing the correlation between each pixel and its neighbouring pixels is essential. Compared with traditional self-attention and convolutional methods, the NA mechanism effectively captures local textures and details, providing higher fidelity in processing local information. This enhanced precision in detecting subtle quality differences makes it especially suitable for the IQA task. Accordingly, we developed a local perceptual module based on the NA mechanism.

Let $Q_k$, $K_k$, and $V_k \in \mathbb{R}^{1 \times l}$ denote the $l$-dimensional query, key, and value vectors of the $k$-th input, respectively, obtained through linear projection. The neighborhood of size $n$ for the $k$-th input is defined as $\mathcal{N}_n(k) = \{\rho_1(k), \ldots, \rho_n(k)\}$, where $\rho_i(k)$ represents the $i$-th nearest neighbor to $k$. The neighbourhood attention operation on $k$ is given by:
\begin{equation}\label{eq:NA}
 N{\hspace{-2pt}A_n}(k) \hspace{-2pt}=\hspace{-2pt} so\hspace{-1pt}ftmax\hspace{-2pt} \left(\hspace{-2pt}{\frac{{{Q_k} \hspace{-2pt}\cdot\hspace{-2pt} {\mathbb{K}_n}(k) \hspace{-2pt}+\hspace{-2pt} {\mathbb{B}_n}(k)}}{{\sqrt d }}} \hspace{-2pt}\right)\hspace{-2pt} \cdot\hspace{-2pt} \mathbb{V}_n^\prime(k),
\end{equation}
where $\mathbb{K}_k(i) = [K_{\rho_1(k)}^\prime, \ldots, K^\prime_{\rho_k(k)}] \in \mathbb{R}^{l \times n}$ is the key matrix, $\mathbb{V}_n(k) = [V_{\rho_1(k)}^\prime, \ldots, V_{\rho_n(k)}^\prime] \in \mathbb{R}^{l \times n} $ is the value matrix, and $\mathbb{B}_n(k) = [b_{(k, \rho_1(k))}, \ldots, b_{(k, \rho_n(k))}] \in \mathbb{R}^{1 \times n}$ is the relative positional bias matrix. $A^\prime$ denotes the transpose of matrix $A$.

\begin{figure}[t]
  \centering
  \includegraphics[width=0.48\textwidth,height=0.25\textwidth]{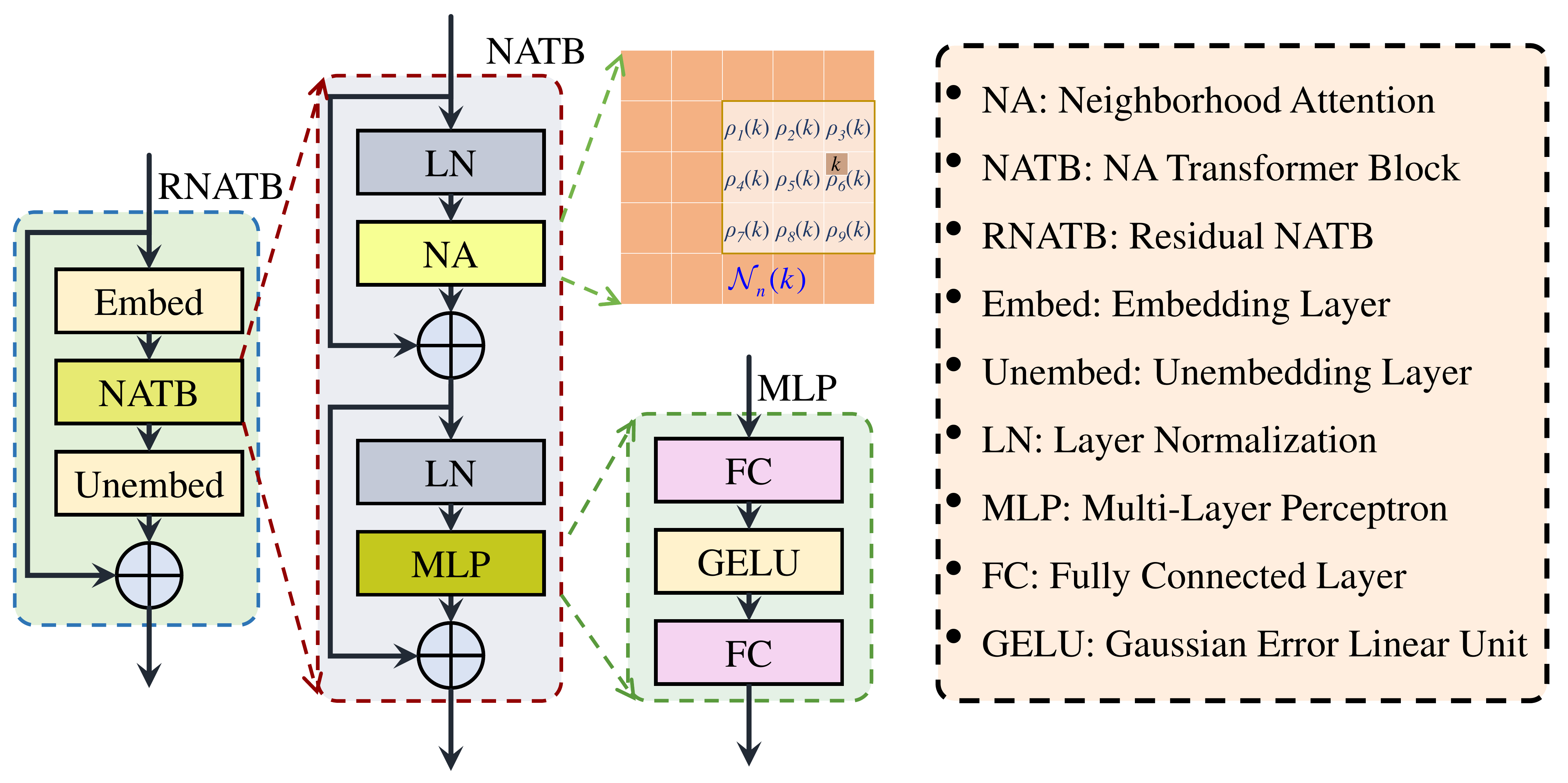}\\
\caption{The structure of the local perceptual module. It utilizes the NA mechanism, a technique that combines local inductive biases with translational invariance, to effectively aggregate pixel features and their surrounding neighborhood. This module enables a detailed analysis of each pixel in relation to its immediate environment.}

\label{fig:encoderdecoder}
\end{figure}

Therefore, as illustrated in \textbf{Fig. \ref{fig:encoderdecoder}}, we define the local distortion-aware functions \(f_1 \) and \(f_2 \) in equation (3) as two distinct residual NA transformer blocks (RNATB):
\begin{gather}\label{eq:f1f2}
f_1(I) = RNATB_1(I)\\
f_2(I) = RNATB_2(I)
\end{gather}

In addition to perceiving local detail distortions, the model must also capture global perceptual features. This capability helps the model understand the overall structure and context of the image. For efficient extraction of global features, we utilize the architecture of the classical CNN model ResNet50 \cite{37}, with modifications to the input channel count. Consequently, we define the global perceptual function in Eq. (\ref{eq:problem}) as follows:
\begin{equation}\label{eq:Qhat}
\begin{aligned}
  {{\hat{Q}}} &= f_3 \big(I, {{f_1}({I})\otimes{d_1}, {f_2}({I})\otimes{d_2})}\big)\\
  &=ResNet \Big(I\oplus\big({{f_1}({I})\otimes{d_1}\big)\oplus\big({f_2}({I})\otimes{d_2}} \big)\Big)
\end{aligned}
\end{equation}
where $\otimes$ denotes the concatenation operation and where $\hat{Q}$ represents the predicted image quality score. This dual approach to feature extraction, encompassing both local and global perspectives, enhances the comprehensiveness and precision of the assessment process, thereby improving the accuracy of the UIQA.

Finally, the optimization of Eq. (\ref{eq:problem}) is equivalent to minimizing the following loss function:
\begin{equation}\label{eq:L}
  \mathop {\min }\limits_{{f_1},{f_2},{f_3}} \mathcal L = \| {Q - \hat Q} \|_2
\end{equation}
where $\| \cdot \|_2$ denotes the $L$2 norm.

\section{Experiments}
\label{sec:exp}

\subsection{Experimental Settings}
\label{subsec:exp_set}

\renewcommand\arraystretch{1}
\begin{table*}[t]
\caption{Performance comparison in terms of the PLCC, SRCC, KRCC and RMSE on the basis of the SAUD2.0 \cite{30} dataset.}
\label{tab:performance1}
\centering
\small
\begin{threeparttable}
\begin{tabular}{C{0.32\columnwidth}|C{0.15\columnwidth}C{0.15\columnwidth}C{0.15\columnwidth}C{0.15\columnwidth}|C{0.15\columnwidth}C{0.15\columnwidth}C{0.15\columnwidth}C{0.15\columnwidth}}
\toprule[0.8pt]
\bottomrule[0.5pt]
\rowcolor[HTML]{DFDFDF} ~ & \multicolumn{4}{c|}{SAUD2.0 (80\%) $\rightarrow$ SAUD2.0 (20\%)}& \multicolumn{4}{c}{SAUD2.0 (100\%) $\rightarrow$ UID2021 (100\%) }\\
\cline{2-5}\cline{6-9}
\rowcolor[HTML]{DFDFDF} \multirow{-2}*{IQA Method} & PLCC & SRCC & KRCC & RMSE & PLCC & SRCC & KRCC & RMSE\\
\hline
BRISQUE \cite{4} & 0.5276 & 0.5018 & 0.3533 & 17.9813 & 0.3248 & 0.3097 & 0.2088 & 2.0384\\
\rowcolor[HTML]{DFDFDF} BLIINDSII \cite{23} & 0.4816 & 0.4449 & 0.3060 & 19.4931 & 0.1635 & 0.1701 & 0.1145 & 2.1262\\
BIQME \cite{32} & 0.4938 & 0.4553 & 0.3168 & 17.6922 & 0.4274 & 0.3652 & 0.2484 & 1.9485\\
\rowcolor[HTML]{DFDFDF} FRIQUEE \cite{33} & 0.7882 & 0.7738 & 0.5843 & 13.1197 & 0.3150 & 0.3042 & 0.2053 & 2.0455\\
NRSL \cite{24} & 0.5176 & 0.4904 & 0.3382 & 18.2603 & 0.2887 & -0.2781 & -0.1867 & 2.0634\\
\rowcolor[HTML]{DFDFDF} RISE \cite{25} & 0.1310 & 0.1037 & 0.0713 & 21.0194 & 0.2760 & 0.2647 & 0.1749 & 2.0715\\
DIIVINE \cite{26} & 0.5151 & 0.5016 & 0.3523 & 18.2658 & 0.1214 & 0.1199 & 0.0798 & 2.1393\\
\rowcolor[HTML]{DFDFDF} DBCNN \cite{27} & 0.7744 & 0.7633 & 0.5704 & 14.6094 & 0.3840 & 0.3794 & 0.2564 & 1.9589\\
WaDIQaM \cite{31} & 0.8565 & 0.8408 & 0.6617 & 10.8477 & 0.5099 & 0.4876 & 0.3414 & 1.8250\\
\rowcolor[HTML]{DFDFDF} DEIQT \cite{50} & \textcolor[rgb]{0.00,0.45,0.00}{$\mathbf{0.9114}$} & \textcolor[rgb]{0.00,0.45,0.00}{$\mathbf{0.9007}$} & \textcolor[rgb]{0.00,0.45,0.00}{$\mathbf{0.7272}$} & \textcolor[rgb]{1.00,0.00,0.00}{$\mathbf{8.5378}$} & \textcolor[rgb]{0.00,0.45,0.00}{$\mathbf{0.5719}$} & \textcolor[rgb]{0.00,0.45,0.00}{$\mathbf{0.5709}$} & \textcolor[rgb]{0.00,0.45,0.00}{$\mathbf{0.4028}$} & \textcolor[rgb]{0.00,0.00,1.00}{$\mathbf{1.8202}$}\\
\hline
B-FEN \cite{28} & 0.8512 & 0.8441 & 0.6595 & 10.8642 & 0.4841 & 0.4766 & 0.3261 & 1.8564\\
\rowcolor[HTML]{DFDFDF} UCIQE \cite{15} & 0.3674 & 0.3719 & 0.2597 & 19.9196 & 0.3238 & 0.3080 & 0.2080 & 2.0391\\
UIQM \cite{16} & 0.3498 & 0.2902 & 0.1993 & 19.9120 & 0.2480 & 0.1395 & 0.0926 & 2.0879\\
\rowcolor[HTML]{DFDFDF} CCF \cite{17} & 0.1664 & 0.1624 & 0.1088 & 20.9302 & 0.3204 & 0.1759 & 0.1164 & 2.0416\\
FDUM \cite{14} & 0.2815 & 0.2613 & 0.1785 & 20.1235 & 0.3229 & 0.3364 & 0.2307 & 2.0397\\
\rowcolor[HTML]{DFDFDF} NUIQ \cite{29} & 0.7413 & 0.7480 & 0.5471 & 14.1152 & 0.3143 & 0.3134 & 0.2078 & 2.0460\\
MCOLE \cite{30} & \textcolor[rgb]{0.00,0.00,1.00}{$\mathbf{0.8838}$} & \textcolor[rgb]{0.00,0.00,1.00}{$\mathbf{0.8748}$} & \textcolor[rgb]{0.00,0.00,1.00}{$\mathbf{0.6887}$} & \textcolor[rgb]{0.00,0.00,1.00}{$\mathbf{9.7132}$} & \textcolor[rgb]{0.00,0.00,1.00}{$\mathbf{0.5520}$} & \textcolor[rgb]{0.00,0.00,1.00}{$\mathbf{0.5340}$} & \textcolor[rgb]{0.00,0.00,1.00}{$\mathbf{0.3779}$} & \textcolor[rgb]{0.00,0.45,0.00}{$\mathbf{1.7690}$}\\
\rowcolor[HTML]{DFDFDF} PIGUIQA (ours) & \textcolor[rgb]{1.00,0.00,0.00}{$\mathbf{0.9142}$} & \textcolor[rgb]{1.00,0.00,0.00}{$\mathbf{0.9043}$} & \textcolor[rgb]{1.00,0.00,0.00}{$\mathbf{0.7333}$} & \textcolor[rgb]{0.00,0.45,0.00}{$\mathbf{8.9984}$} & \textcolor[rgb]{1.00,0.00,0.00}{$\mathbf{0.5953}$} & \textcolor[rgb]{1.00,0.00,0.00}{$\mathbf{0.5831}$} & \textcolor[rgb]{1.00,0.00,0.00}{$\mathbf{0.4145}$} & \textcolor[rgb]{1.00,0.00,0.00}{$\mathbf{1.7342}$} \\
\toprule[0.5pt]
\bottomrule[0.8pt]
\end{tabular}
\begin{tablenotes}
  \footnotesize
  \item ``A ($a\%$) $\rightarrow$ B ($b\%$)'' means training on the $a\%$ data of dataset ``A'' and testing on the $b\%$ data of dataset ``B''. The general-purpose methods and the specific-purpose methods are separated by a horizontal line. In each column, the three best values are bolded in \textcolor[rgb]{1.00,0.00,0.00}{\textbf{red (1st)}}, \textcolor[rgb]{0.00,0.45,0.00}{\textbf{green (2nd)}}, and \textcolor[rgb]{0.00,0.00,1.00}{\textbf{blue (3rd)}} colors, respectively. Some of the data in the table are obtained from the literature.
\end{tablenotes}
\end{threeparttable}
\end{table*}

\begin{figure*}[t]
\small
\begin{minipage}{0.245\linewidth}
  \centering
  \centerline{\includegraphics[width=0.2\paperwidth, height=0.19\paperwidth]{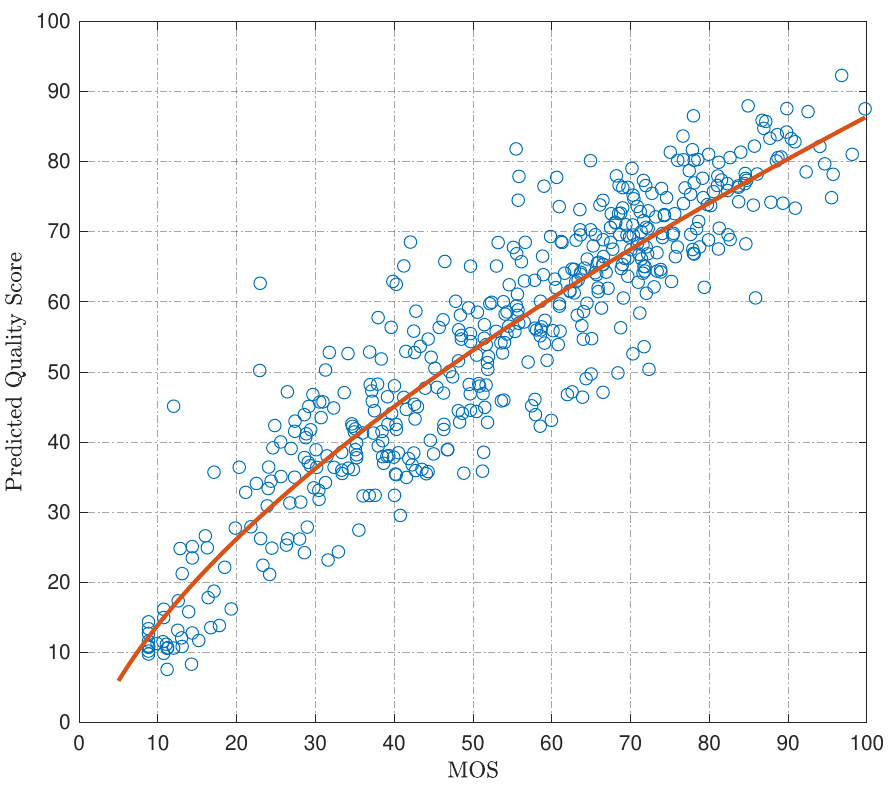}}
  \centerline{\scriptsize{(a) Train and test on SAUD2.0}}
\end{minipage}
\hfill
\begin{minipage}{0.245\linewidth}
  \centering
  \centerline{\includegraphics[width=0.2\paperwidth, height=0.19\paperwidth]{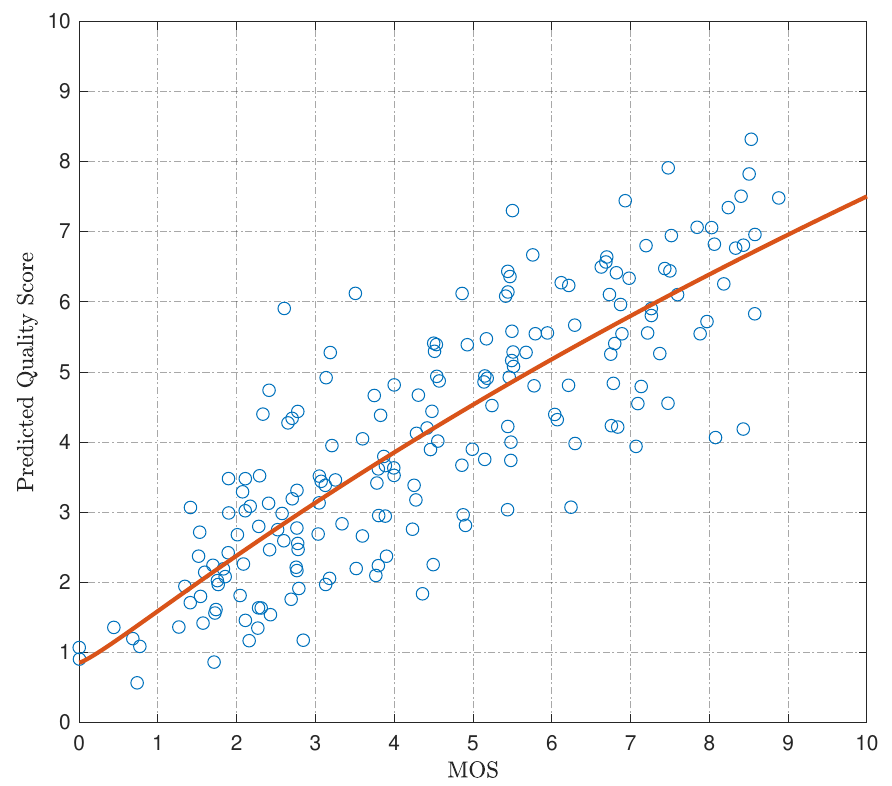}}
  \centerline{\scriptsize{(b) Train and test on UID2021}}
\end{minipage}
\hfill
\begin{minipage}{0.245\linewidth}
  \centering
  \centerline{\includegraphics[width=0.2\paperwidth, height=0.19\paperwidth]{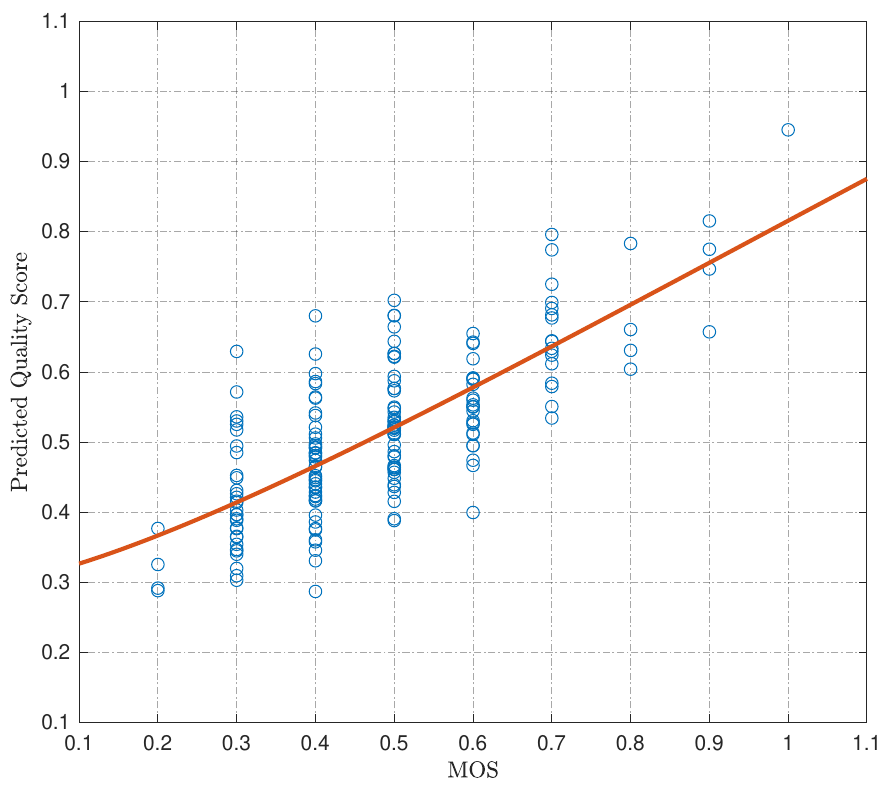}}
  \centerline{\scriptsize{(c) Train and test on UWIQA}}
\end{minipage}
\hfill
\begin{minipage}{0.245\linewidth}
  \centering
  \centerline{\includegraphics[width=0.2\paperwidth, height=0.19\paperwidth]{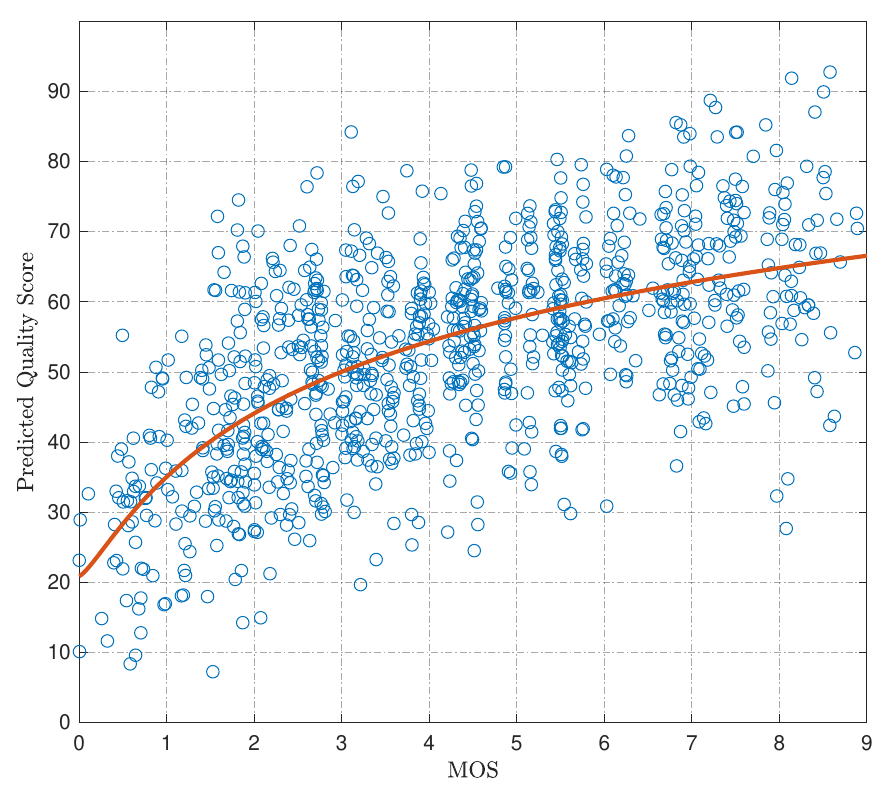}}
  \centerline{\scriptsize{(d) Train on SAUD2.0 and test on UID2021}}
\end{minipage}

\caption{Scatter plots of the quality scores predicted by the proposed model against the MOS. The red curves are the 5PL fitting functions.}
\label{fig:result_fig1}
\end{figure*}

\begin{table}[t]
\caption{Performance comparison in terms of the PLCC, SRCC, KRCC and RMSE on the UID2021 dataset \cite{22}.}
\label{tab:performance2}
\centering
\small
\begin{threeparttable}
\begin{tabular}{C{0.29\columnwidth}|C{0.125\columnwidth}C{0.125\columnwidth}C{0.125\columnwidth}C{0.125\columnwidth}}
\toprule[0.8pt]
\bottomrule[0.5pt]
\rowcolor[HTML]{DFDFDF} IQA Method & PLCC & SRCC & KRCC & RMSE \\
\hline
BRISQUE \cite{4} & 0.6439 & 0.6343 & 0.4623 & 1.6407 \\
\rowcolor[HTML]{DFDFDF} BLIINDSII \cite{23} & 0.5451 & 0.5216 & 0.3688 & 1.7934 \\
NIQE \cite{6} & 0.3384 & 0.3304 & 0.2219 & 2.0464 \\
\rowcolor[HTML]{DFDFDF} IL-NIQE \cite{7} & 0.4644 & 0.4630 & 0.4321 & 1.9121 \\
NRSL \cite{24} & 0.6643 & 0.6504 & 0.4655 & 1.6016 \\
\rowcolor[HTML]{DFDFDF} RISE \cite{25} & 0.6219 & 0.6034 & 0.4314 & 1.6812 \\
DIIVINE \cite{26} & 0.6264 & 0.6112 & 0.4363 & 1.6716 \\
\rowcolor[HTML]{DFDFDF} DBCNN \cite{27} & 0.7594 & 0.7535 & 0.5523 & 1.3862 \\
WaDIQaM \cite{31} & \textcolor[rgb]{0.00,0.00,1.00}{$\mathbf{0.7736}$} & 0.7659 & \textcolor[rgb]{0.00,0.00,1.00}{$\mathbf{0.5750}$} & \textcolor[rgb]{0.00,0.00,1.00}{$\mathbf{1.3592}$} \\
\hline
\rowcolor[HTML]{DFDFDF} B-FEN \cite{28} & 0.7713 & \textcolor[rgb]{0.00,0.00,1.00}{$\mathbf{0.7674}$} & 0.5732 & 1.3696 \\
UCIQE \cite{15} & 0.6474 & 0.6150 & 0.4503 & 1.6335 \\
\rowcolor[HTML]{DFDFDF} UIQM \cite{16} & 0.4760 & 0.4613 & 0.3192 & 1.8769 \\
CCF \cite{17} & 0.5208 & 0.4236 & 0.2990 & 1.8351 \\
\rowcolor[HTML]{DFDFDF} FDUM \cite{14} & 0.6092 & 0.5823 & 0.4189 & 1.7057 \\
NUIQ \cite{29} & 0.7266 & 0.7168 & 0.5293 & 1.4762 \\
\rowcolor[HTML]{DFDFDF} MCOLE \cite{30} & \textcolor[rgb]{0.00,0.45,0.00}{$\mathbf{0.7977}$} & \textcolor[rgb]{0.00,0.45,0.00}{$\mathbf{0.7915}$} & \textcolor[rgb]{0.00,0.45,0.00}{$\mathbf{0.6024}$} & \textcolor[rgb]{1.00,0.00,0.00}{$\mathbf{1.2964}$} \\
PIGUIQA (ours) & \textcolor[rgb]{1.00,0.00,0.00}{$\mathbf{0.8162}$} & \textcolor[rgb]{1.00,0.00,0.00}{$\mathbf{0.8146}$} & \textcolor[rgb]{1.00,0.00,0.00}{$\mathbf{0.6225}$} & \textcolor[rgb]{0.00,0.45,0.00}{$\mathbf{1.3269}$} \\
\toprule[0.5pt]
\bottomrule[0.8pt]
\end{tabular}
\begin{tablenotes}
  \footnotesize
  \item The general-purpose methods and the specific-purpose methods are separated by a horizontal line. In each column, the three best values are bolded in \textcolor[rgb]{1.00,0.00,0.00}{\textbf{red (1st)}}, \textcolor[rgb]{0.00,0.45,0.00}{\textbf{green (2nd)}}, and \textcolor[rgb]{0.00,0.00,1.00}{\textbf{blue (3rd)}} colors, respectively. Some of the data in the table are obtained from the literature.
\end{tablenotes}
\end{threeparttable}
\end{table}

\textbf{Databases.} The experiments are conducted on three databases: SAUD2.0 (2024) \cite{30}, UID2021 (2023) \cite{22}, and UWIQA (2021) \cite{3}. The images in these datasets are sourced from real-world scenarios, and their subjective scores are MOSs, which are obtained through the single-stimulus absolute category rating (SS-ACR) methodology. Specifically, SAUD2.0 comprises 2,600 images (200 raw and 2,400 enhanced), UID2021 consists of 1,060 images (60 raw and 1,000 enhanced), and UWIQA includes 890 raw images.

\textbf{Implementation details.} The proposed FIGUIQA network is trained
\footnote{The implementation code is programmed in PyTorch and is publicly available at \href{https://anonymous.4open.science/r/PIGUIQA-A465/}{https://anonymous.4open.science/r/PIGUIQA-A465/}. The experiments are run on a PC with a Linux operating system and are configured with an NVIDIA GeForce RTX 4090 GPU with 24 GB of memory.} using the default settings of the Adam optimizer. The learning rate is set to $10^{-4}$. To ensure the stability of the final model, a total of 1000 epochs are conducted during training. To prevent overfitting, data augmentation techniques such as random horizontal and vertical flipping, as well as rotation within the range of $[-15,15]$ degrees, are employed. The size of patches is set to $16 \times 16$, which is a common setting in computer vision. SyreaNet \cite{1} is selected as the pre-trained underwater imaging estimation model in the framework based on its state-of-the-art performance.

\textbf{Evaluation criteria.} To validate the effectiveness and performance of the proposed UIQA method, we employ four commonly used evaluation metrics: the Pearson linear correlation coefficient (PLCC), the Spearman rank correlation coefficient (SRCC), the Kendall rank correlation coefficient (KRCC), and the root mean square error (RMSE). In accordance with standard practice, we apply nonlinear five-parameter logistic regression (5PL) to fit the relationship between the predicted scores and the MOS prior to calculating the PLCC and RMSE.

\subsection{Performance Comparison}
\label{subsec:performance}

We conducted a systematic comparative analysis with 28 commonly-used or state-of-the-art no-reference image quality assessment (NR-IQA) and UIQA methods, including BRISQUE \cite{4}, BLIINDSII \cite{23}, NFERM \cite{5}, NIQE \cite{6}, IL-NIQE \cite{7}, SNP-NIQE \cite{8}, BIQME \cite{32}, FRIQUEE \cite{33}, PIQE \cite{9}, NRSL \cite{24}, RISE \cite{25}, DIIVINE \cite{26}, NPQI \cite{10}, dipIQ \cite{11}, HyperIQA \cite{12}, WaDIQaM \cite{31}, DEIQT \cite{50}, UNIQUE \cite{13}, DBCNN \cite{27}, B-FEN \cite{28}, UCIQE \cite{15}, UIQM \cite{16}, CCF \cite{17}, FDUM \cite{14}, NUIQ \cite{29}, MCOLE \cite{30}, Twice-Mix \cite{42}, and UIQI \cite{3}. To ensure the reliability and stability of the proposed NAFRAD method, we repeat the training and testing process 10 times with randomized data partitioning, using the average of the test results as the final metric.

\subsubsection{\textbf{Single-Dataset Scenarios}}
\label{subsubsec:s-evaluation}

In the experimental scenario involving single datasets, we randomly selected 80\% of the data from each dataset for training and used the remaining 20\% for evaluation. The experimental results are presented in \textbf{Tables \ref{tab:performance1}}, \textbf{\ref{tab:performance2}}, and \textbf{\ref{tab:performance3}}. Our proposed method achieves the highest performance on the SAUD2.0 dataset, significantly surpassing the second-place method, MCOLE \cite{30}, demonstrating substantial performance improvement. Specifically, our image quality assessment method ranked first on the SAUD2.0 dataset and nearly achieved first place on the SUID2021 dataset, with only a slight margin separating it from the top performance. For the UWIQA dataset, our method ranks second. Overall, there is a noticeable trend of performance improvement as the dataset size increases, moving from UWIQA to SUID2021 and then to SAUD2.0. This trend in performance enhancement is closely related to the dataset size, which is attributable to the use of deep learning techniques in our proposed method. The effectiveness of deep learning models is typically influenced by the availability of large quantities of high-quality data.

\textbf{Fig. \ref{fig:result_fig1}} shows scatter plots of the testing dataset. This figure clearly shows that the quality scores predicted by the proposed PIGUIQA model are highly correlated with the actual MOS, with relatively small prediction errors. This indicates that our method achieves a high level of accuracy in quality score prediction. A closer examination of \textbf{Fig. \ref{fig:result_fig1}}(c) reveals some unique characteristics of the UWIQA dataset. UWIQA is relatively small and has low quantization precision in its MOS labels, which are broadly categorized into only ten levels. This low-precision labelling poses challenges for training deep learning models, resulting in less effective learning of subtle image quality differences compared with other datasets.

\subsubsection{\textbf{Cross-Dataset Scenarios}}
\label{subsubsec:c-evaluation}

To evaluate the generalizability of the proposed PIGUIQA method, we conducted cross-dataset experiments on the SAUD2.0 and UID2021 datasets. In this experiment, we initially trained the model on the SAUD2.0 dataset and subsequently applied the trained model to the UID2021 dataset for testing. The experimental results, presented in \textbf{Table \ref{tab:performance1}} and \textbf{Fig. \ref{fig:result_fig1}}(d), demonstrate that our method exhibits exceptional performance under cross-dataset conditions, further validating its robustness and strong generalization ability across different datasets.

The remarkable performance in cross-dataset evaluations can be attributed to our method's careful incorporation of prior knowledge related to underwater imaging models during the design process. The characteristics of underwater images are influenced by various factors, such as water turbidity, lighting conditions, and image distortion, resulting in significant domain specificity. To effectively address these characteristics, our approach integrates physical prior knowledge of the underwater imaging process into the modelling framework. This allows the model to adapt successfully to various types of underwater images, even when faced with different datasets.

\subsection{Ablation Experiments}
\label{subsec:ablation}

To validate the effectiveness of each module in the PIGUIQA method, we conducted ablation experiments on the SAUD2.0 dataset. In this experiment, we established two variables: the perception of transmission attenuation distortion and the consideration of backscattering distortion. The results, as shown in \textbf{Table \ref{tab:performance4}}, indicate that when we remove one of the modules (i.e., retain only one distortion perception module), the model's performance slightly decreases. This suggests that while each module independently contributes to the model's ability to perceive certain aspects of image quality, they are not entirely independent and exhibit some interdependence. However, when both modules are removed simultaneously, there is a significant drop in the model's performance, particularly in terms of accuracy and robustness in image quality assessment. This finding underscores the critical role that transmission attenuation distortion and backscattering distortion play as key factors in the underwater imaging process, significantly impacting the evaluation of image quality.

\renewcommand\arraystretch{1}
\begin{table}[t]
\caption{Performance comparison in terms of PLCC, SRCC, KRCC and RMSE on the UWIQA dataset \cite{14}.}
\label{tab:performance3}
\centering
\small
\begin{threeparttable}
\begin{tabular}{C{0.29\columnwidth}|C{0.125\columnwidth}C{0.125\columnwidth}C{0.125\columnwidth}C{0.125\columnwidth}}
\toprule[0.8pt]\bottomrule[0.5pt]
\rowcolor[HTML]{DFDFDF} IQA Method & PLCC & SRCC & KRCC & RMSE \\\hline
BRISQUE \cite{4} & 0.3669 & 0.3456 & 0.2562 & 0.1415 \\\rowcolor[HTML]{DFDFDF}
NFERM \cite{5} & 0.3925 & 0.3486 & 0.2595 & 0.1398 \\
NIQE \cite{6} & 0.4687 & 0.4347 & 0.3243 & 0.1343 \\\rowcolor[HTML]{DFDFDF}
IL-NIQE \cite{7} & 0.4421 & 0.4686 & 0.3476 & 0.1364 \\
SNP-NIQE \cite{8} & 0.5897 & 0.5516 & 0.4199 & 0.1228 \\\rowcolor[HTML]{DFDFDF}
PIQE \cite{9} & 0.3224 & 0.2084 & 0.1492 & 0.1441 \\
NPQI \cite{10} & 0.6361 & 0.6078 & 0.4667 & 0.1173 \\\rowcolor[HTML]{DFDFDF}
dipIQ \cite{11} & 0.1369 & 0.0869 & 0.0641 & 0.1506 \\
HyperIQA \cite{12} & \textcolor[rgb]{0.00,0.00,1.00}{$\mathbf{0.6799}$} & 0.6501 & 0.5040 & \textcolor[rgb]{0.00,0.00,1.00}{$\mathbf{0.1114}$} \\\rowcolor[HTML]{DFDFDF}
UNIQUE \cite{13} & 0.2386 & 0.2496 & 0.1835 & 0.1476 \\\hline
UCIQE \cite{15} & 0.6261 & 0.6271 & 0.4863 & 0.1185 \\\rowcolor[HTML]{DFDFDF}
UIQM \cite{16} & 0.5928 & 0.5960 & 0.4563 & 0.1225 \\
CCF \cite{17} & 0.4634 & 0.4456 & 0.3344 & 0.1348 \\\rowcolor[HTML]{DFDFDF}
FDUM \cite{14} & 0.6462 & \textcolor[rgb]{0.00,0.00,1.00}{$\mathbf{0.6780}$} & \textcolor[rgb]{0.00,0.00,1.00}{$\mathbf{0.5289}$} & 0.1160 \\
Twice-Mix \cite{42} & 0.4422 & 0.4727 & 0.3501 & 0.1289 \\\rowcolor[HTML]{DFDFDF}
UIQI \cite{3} & \textcolor[rgb]{0.00,0.45,0.00}{$\mathbf{0.7412}$} & \textcolor[rgb]{1.00,0.00,0.00}{$\mathbf{0.7423}$} & \textcolor[rgb]{1.00,0.00,0.00}{$\mathbf{0.5912}$} & \textcolor[rgb]{1.00,0.00,0.00}{$\mathbf{0.1020}$} \\
PIGUIQA (ours) & \textcolor[rgb]{1.00,0.00,0.00}{$\mathbf{0.7476}$} & \textcolor[rgb]{0.00,0.45,0.00}{$\mathbf{0.7149}$} & \textcolor[rgb]{0.00,0.45,0.00}{$\mathbf{0.5726}$} & \textcolor[rgb]{0.00,0.45,0.00}{$\mathbf{0.1083}$} \\\toprule[0.5pt]\bottomrule[0.8pt]
\end{tabular}
\begin{tablenotes}
  \footnotesize
  \item The general-purpose methods and the specific-purpose methods are separated by a horizontal line. In each column, the three best values are bolded in \textcolor[rgb]{1.00,0.00,0.00}{\textbf{red (1st)}}, \textcolor[rgb]{0.00,0.45,0.00}{\textbf{green (2nd)}}, and \textcolor[rgb]{0.00,0.00,1.00}{\textbf{blue (3rd)}} colors, respectively. Some of the data in the table are obtained from the literature.
\end{tablenotes}
\end{threeparttable}
\end{table}

\begin{table}[t]
\caption{Ablation experimental results in terms of the PLCC, SRCC, KRCC and RMSE on the SAUD2.0 dataset \cite{30}.}
\label{tab:performance4}
\centering
\small
\begin{threeparttable}
\begin{tabular}{C{0.29\columnwidth}|C{0.125\columnwidth}C{0.125\columnwidth}C{0.125\columnwidth}C{0.125\columnwidth}}
\toprule[0.8pt]\bottomrule[0.5pt]
\rowcolor[HTML]{DFDFDF} Components & PLCC & SRCC & KRCC & RMSE \\\hline
w/$f_1, d_1, f_2, d_2$ & 0.9142 & 0.9043 & 0.7333 & 8.9984 \\
\rowcolor[HTML]{DFDFDF} w/o $f_1, d_1$ & 0.8840 & 0.8792 & 0.6961 & 10.37 \\
w/o $f_2, d_2$ & 0.8738 & 0.8608 & 0.6862 & 11.19 \\
\rowcolor[HTML]{DFDFDF} w/o $f_1, d_1, f_2, d_2$ & 0.8256 & 0.8320 & 0.6252 & 13.35 \\\toprule[0.5pt]\bottomrule[0.8pt]
\end{tabular}
\end{threeparttable}
\end{table}

\subsection{Visualization of Correlation Performance}

\begin{figure*}[t]

\begin{minipage}{0.195\linewidth}
  \centering
\centerline{\includegraphics[width=1\linewidth, height=1\linewidth]{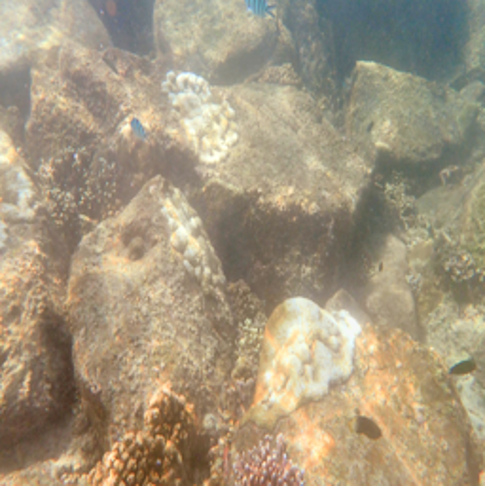}}
\centerline{\scriptsize{(a) MOS:34.97}}
\vspace{4pt}
\end{minipage}
\hfill
\begin{minipage}{0.195\linewidth}
  \centering
\centerline{\includegraphics[width=1\linewidth, height=1\linewidth]{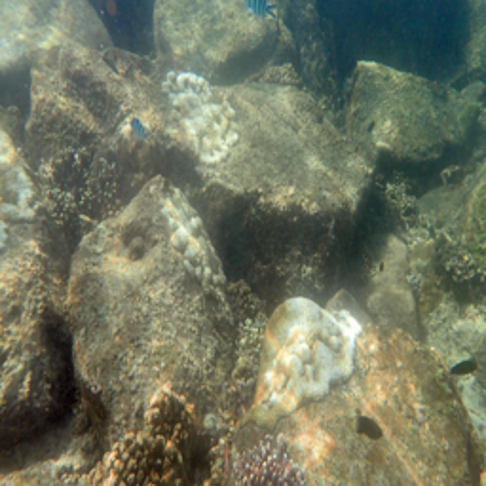}}
\centerline{\scriptsize{(b) MOS:37.06}}
\vspace{4pt}
\end{minipage}
\hfill
\begin{minipage}{0.195\linewidth}
  \centering
\centerline{\includegraphics[width=1\linewidth, height=1\linewidth]{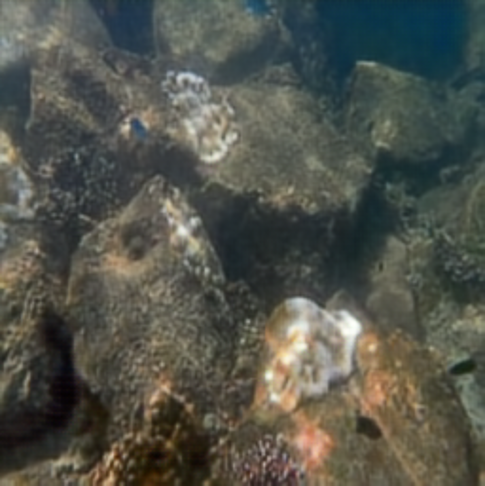}}
\centerline{\scriptsize{(c) MOS:45.75}}
\vspace{4pt}
\end{minipage}
\hfill
\begin{minipage}{0.195\linewidth}
  \centering
\centerline{\includegraphics[width=1\linewidth, height=1\linewidth]{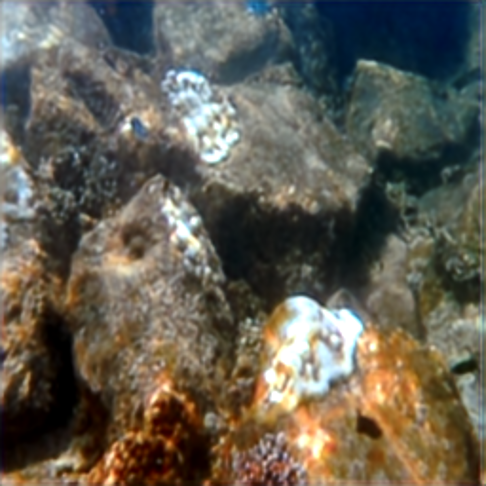}}
\centerline{\scriptsize{(d) MOS:56.03}}
\vspace{4pt}
\end{minipage}
\hfill
\begin{minipage}{0.195\linewidth}
  \centering
\centerline{\includegraphics[width=1\linewidth, height=1\linewidth]{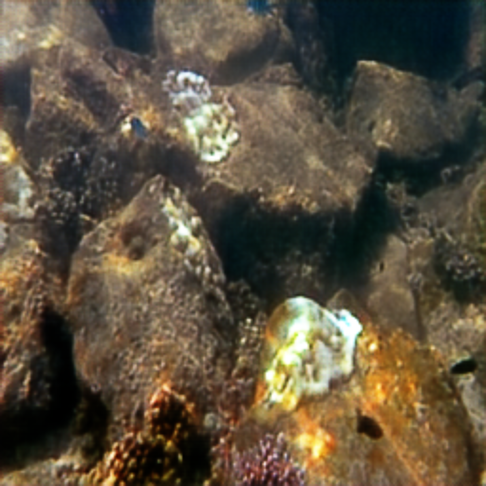}}
\centerline{\scriptsize{(e) MOS:59.59}}
\vspace{4pt}
\end{minipage}

\begin{minipage}{0.195\linewidth}
  \centering
\centerline{\includegraphics[width=1\linewidth, height=1\linewidth]{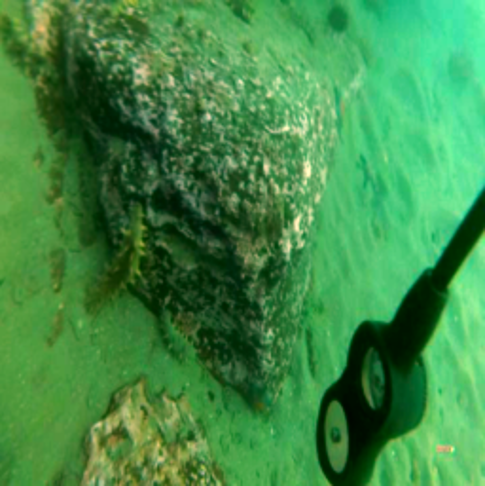}}
\centerline{\scriptsize{(f) MOS:38.15}}
\vspace{4pt}
\end{minipage}
\hfill
\begin{minipage}{0.195\linewidth}
  \centering
\centerline{\includegraphics[width=1\linewidth, height=1\linewidth]{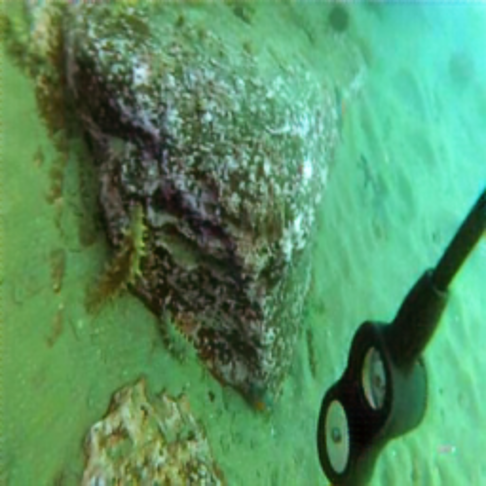}}
\centerline{\scriptsize{(g) MOS:38.28}}
\vspace{4pt}
\end{minipage}
\hfill
\begin{minipage}{0.195\linewidth}
  \centering
\centerline{\includegraphics[width=1\linewidth, height=1\linewidth]{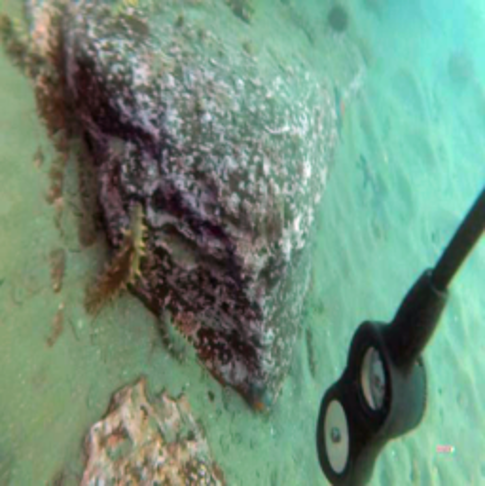}}
\centerline{\scriptsize{(h) MOS:51.24}}
\vspace{4pt}
\end{minipage}
\hfill
\begin{minipage}{0.195\linewidth}
  \centering
\centerline{\includegraphics[width=1\linewidth, height=1\linewidth]{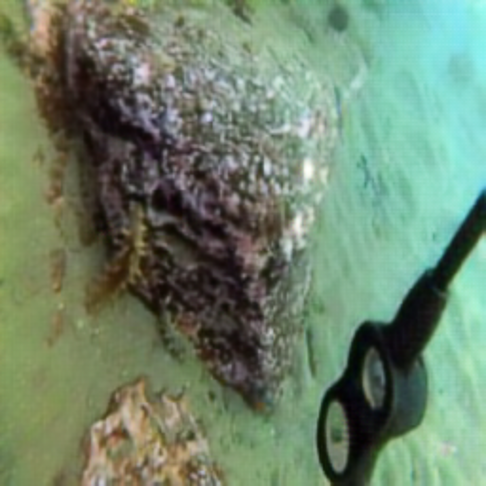}}
\centerline{\scriptsize{(i) MOS:56.89}}
\vspace{4pt}
\end{minipage}
\hfill
\begin{minipage}{0.195\linewidth}
  \centering
\centerline{\includegraphics[width=1\linewidth, height=1\linewidth]{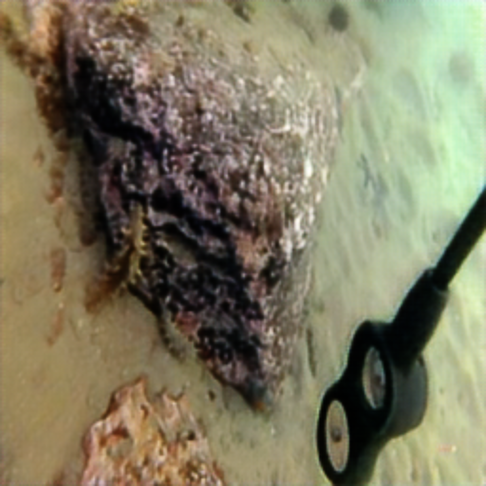}}
\centerline{\scriptsize{(j) MOS:65.01}}
\vspace{4pt}
\end{minipage}

\begin{minipage}{0.195\linewidth}
  \centering
\centerline{\includegraphics[width=1\linewidth, height=1\linewidth]{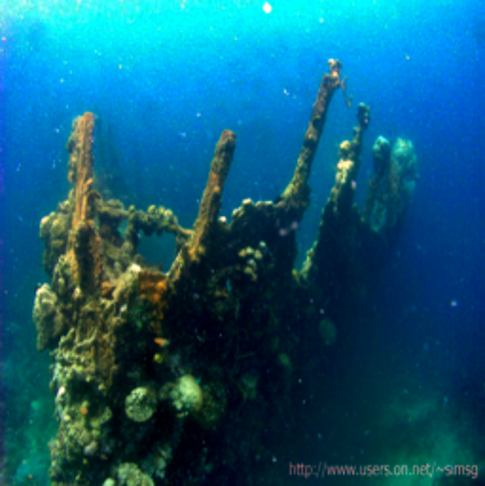}}
\centerline{\scriptsize{(k) MOS:34.97}}
\vspace{6pt}
\end{minipage}
\hfill
\begin{minipage}{0.195\linewidth}
  \centering
\centerline{\includegraphics[width=1\linewidth, height=1\linewidth]{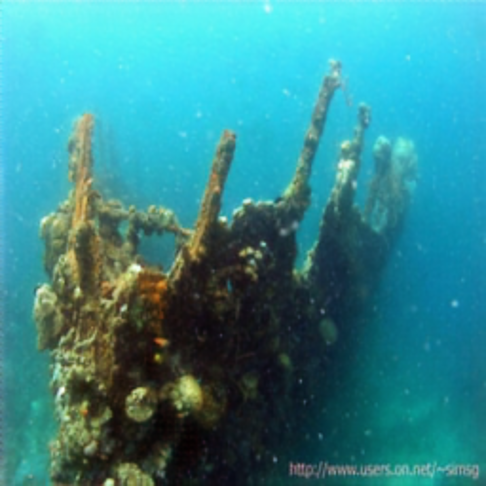}}
\centerline{\scriptsize{(l) MOS:37.06}}
\vspace{6pt}
\end{minipage}
\hfill
\begin{minipage}{0.195\linewidth}
  \centering
\centerline{\includegraphics[width=1\linewidth, height=1\linewidth]{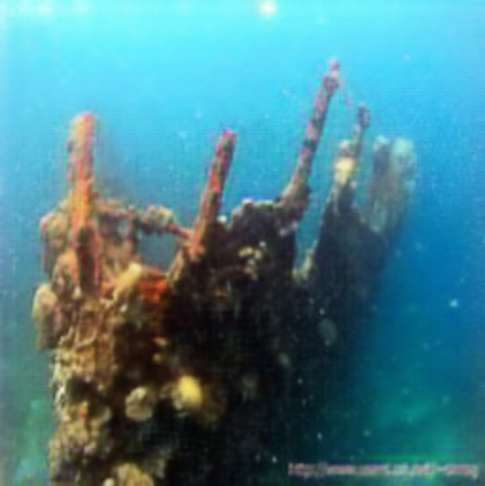}}
\centerline{\scriptsize{(m) MOS:45.75}}
\vspace{6pt}
\end{minipage}
\hfill
\begin{minipage}{0.195\linewidth}
  \centering
\centerline{\includegraphics[width=1\linewidth, height=1\linewidth]{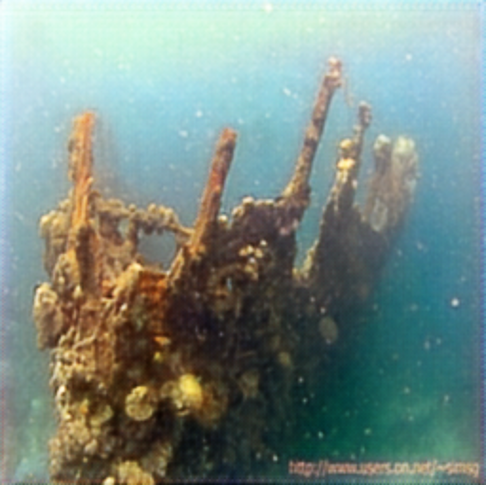}}
\centerline{\scriptsize{(n) MOS:56.03}}
\vspace{6pt}
\end{minipage}
\hfill
\begin{minipage}{0.195\linewidth}
  \centering
\centerline{\includegraphics[width=1\linewidth, height=1\linewidth]{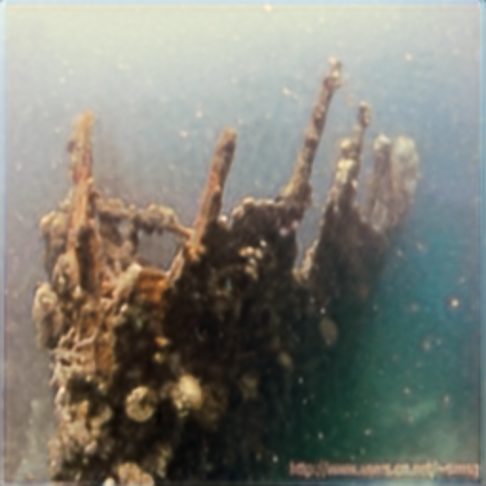}}
\centerline{\scriptsize{(o) MOS:59.59}}
\vspace{6pt}
\end{minipage}

\begin{minipage}{0.33\linewidth}
  \centering
\centerline{\includegraphics[width=0.271\paperwidth, height=0.27\paperwidth]{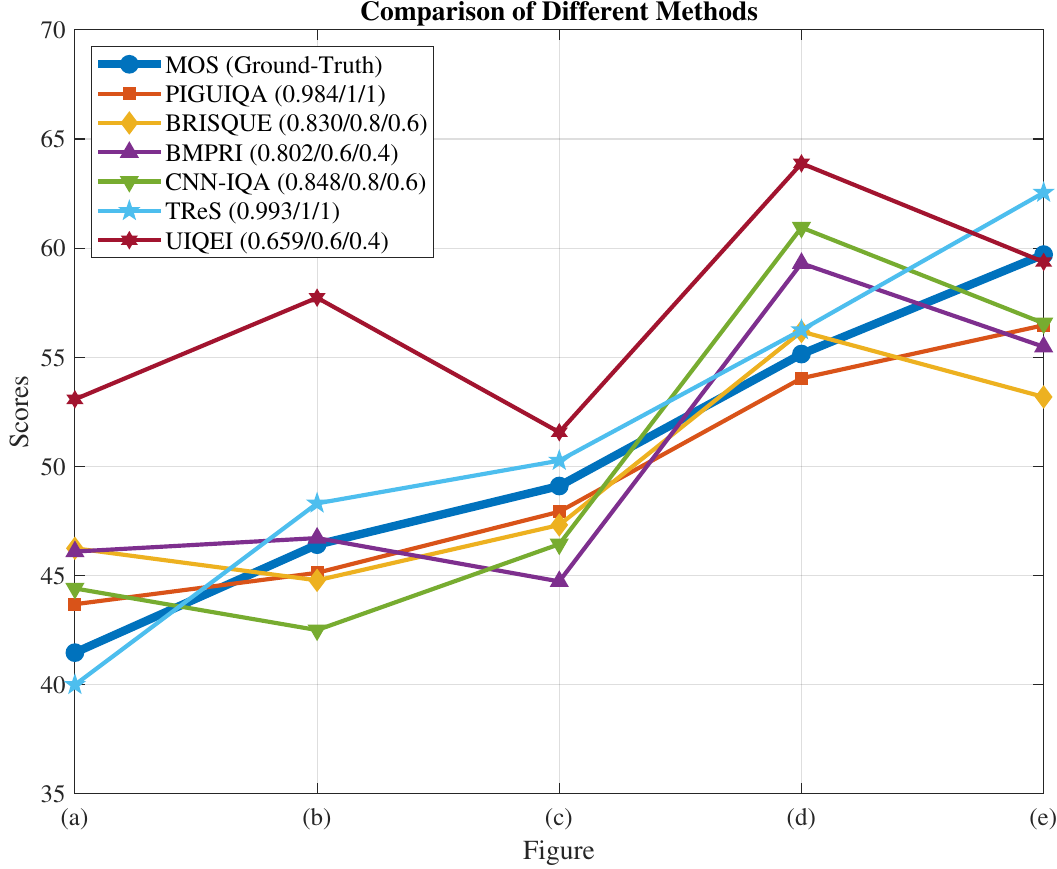}}
\vspace{-4pt}
\centerline{\scriptsize{(p) predictions of (a)-(e)}}
\end{minipage}
\begin{minipage}{0.33\linewidth}
  \centering
\centerline{\includegraphics[width=0.271\paperwidth, height=0.27\paperwidth]{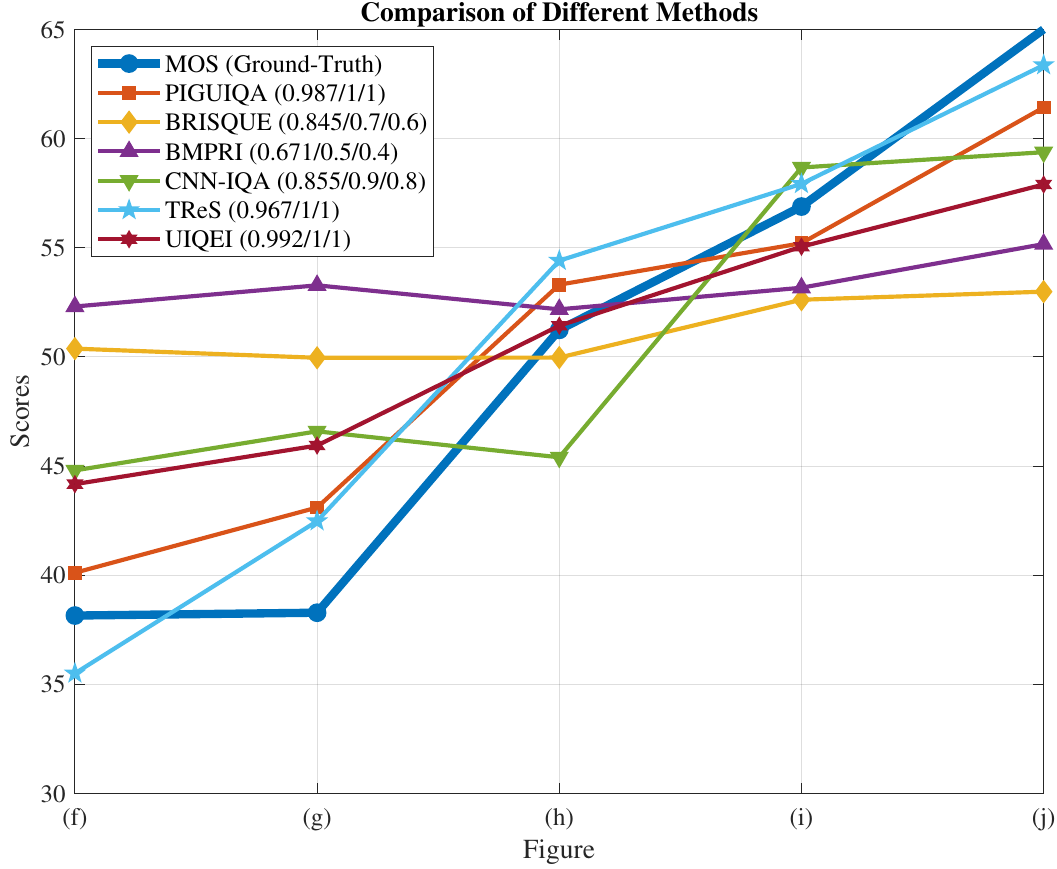}}
\vspace{-4pt}
\centerline{\scriptsize{(q) predictions of (f)-(j)}}
\end{minipage}
\begin{minipage}{0.33\linewidth}
  \centering
\centerline{\includegraphics[width=0.271\paperwidth, height=0.27\paperwidth]{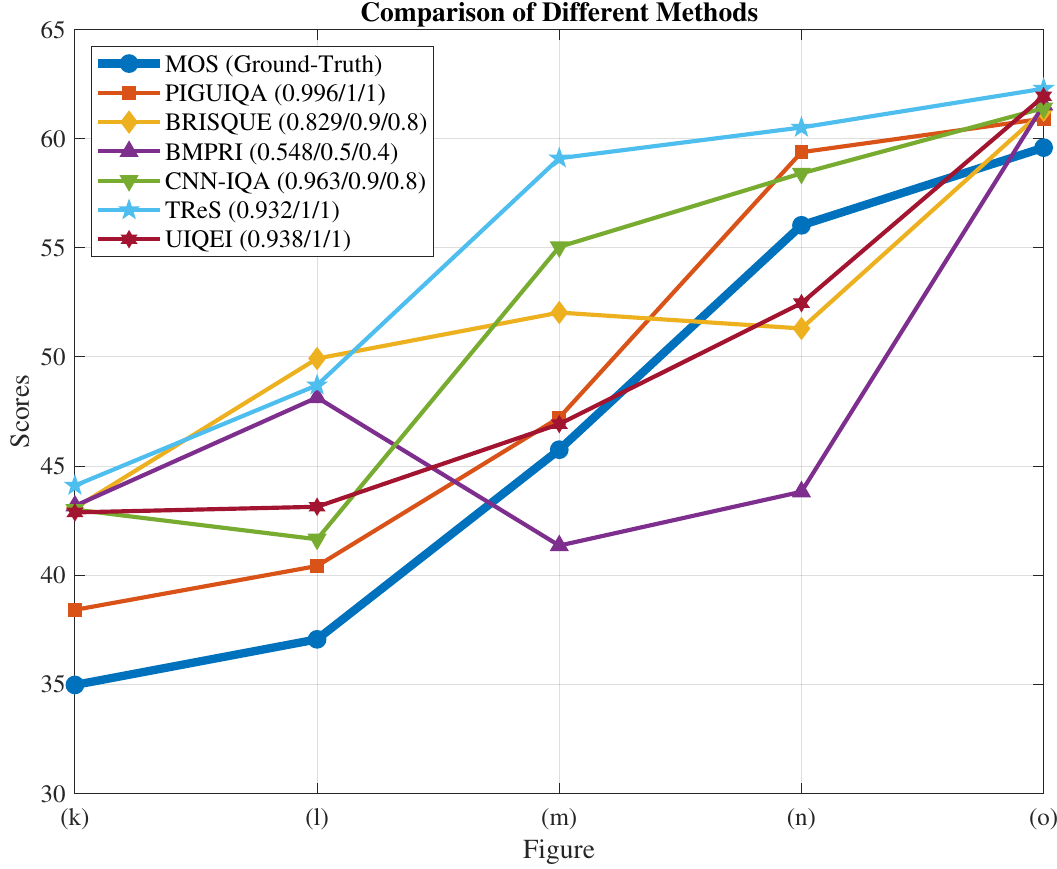}}
\vspace{-4pt}
\centerline{\scriptsize{(r) predictions of (k)-(o)}}
\end{minipage}

\caption{Visualization of correlation performance. In rows 1 to 3, the quality of the images increases progressively from left to right. Subfigures (p), (q), and (r) demonstrate a consistent trend between the proposed method's predicted scores and the mean opinion score (MOS), with the predicted scores closely aligning with the corresponding MOS line. The quantitative results are presented in parentheses within the figure legend as follows: (PLCC/SRCC/KRCC).}
\label{fig:performance}

\end{figure*}

This study proposes a physically imaging-guided framework for underwater image quality assessment (PIGUIQA). To comprehensively evaluate the performance of PIGUIQA, we conducted a series of experiments and compared the results with those of existing state-of-the-art methods.

To visually illustrate the effectiveness of the proposed PIGUIQA method, we select six commonly used and state-of-the-art approaches for comparison. These include traditional general-purpose no-reference image quality assessment (NR-IQA) methods such as BRISQUE \cite{4} and BMPRI \cite{46}; deep learning-based NR-IQA methods such as CNN-IQA \cite{47} and TReS \cite{45}; underwater-specific NR-IQA methods such as UIQEI \cite{48}; and the proposed PIGUIQA. A representative set of image samples was carefully chosen for this analysis. Figure 1 presents these samples along with their corresponding evaluation results, which include predicted scores from different methods and the associated mean opinion score (MOS) values.

For ease of comparison, the images in \textbf{Fig. \ref{fig:performance}} are arranged from left to right on the basis of their MOS values, reflecting a trend in perceived quality from low to high. Images positioned towards the left exhibit lower overall visual quality, primarily due to insufficient enhancement, severe color distortion, and low contrast. As enhancement improves, the perceived quality of underwater images increases, alleviating issues such as color shifts and low contrast, while local details and texture structures become clearer and more discernible.

An analysis of the evaluation results presented in \textbf{Fig. \ref{fig:performance}} yields several important observations:
\begin{itemize}
  \item PIGUIQA consistently provides accurate objective rankings for enhanced images of the same scene, demonstrating its exceptional discriminative ability.
  \item Although TReS also offers correct rankings, the scores produced by PIGUIQA are generally closer to the MOS values, indicating that PIGUIQA not only effectively distinguishes between varying enhancement qualities but also quantifies these differences more accurately.
  \item For underwater images with varying degrees of enhancement from the same scene, most other methods, particularly BMPRI and BRISQUE, exhibit a broader range of predicted quality scores, making it difficult to capture subtle changes and trends. This limitation highlights the challenges these methods face in addressing the unique distortion types present in underwater images.
  \item While TReS and UIQEI have relatively good predictive performance, they still do not match the accuracy of PIGUIQA, underscoring the importance of assessment methods specifically designed for underwater images.
\end{itemize}

While TReS and UIQEI have relatively good predictive performance, they still do not match the accuracy of PIGUIQA, underscoring the importance of assessment methods specifically designed for underwater images.

\begin{figure*}[t]

\begin{minipage}{0.24\linewidth}
  \centering
\centerline{\includegraphics[width=1\linewidth, height=0.8\linewidth]{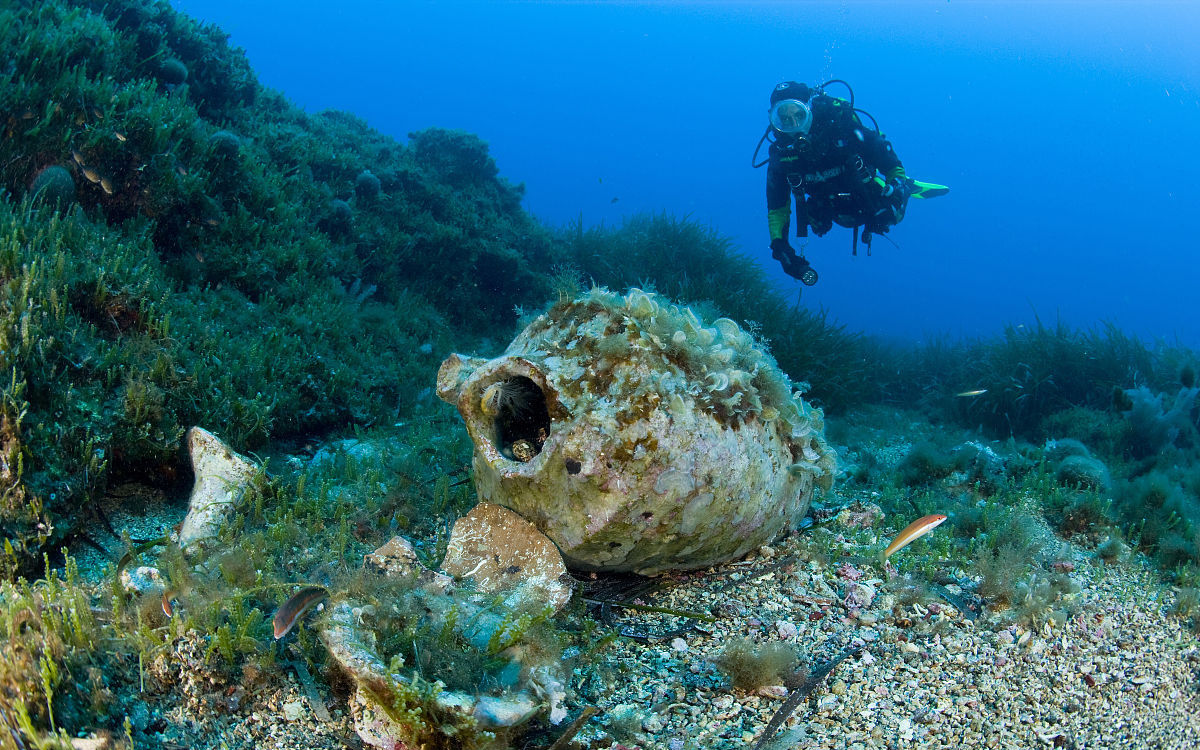}}
\centerline{\scriptsize{(a) MOS:0.9}}
\vspace{4pt}
\end{minipage}
\hfill
\begin{minipage}{0.24\linewidth}
  \centering
\centerline{\includegraphics[width=1\linewidth, height=0.8\linewidth]{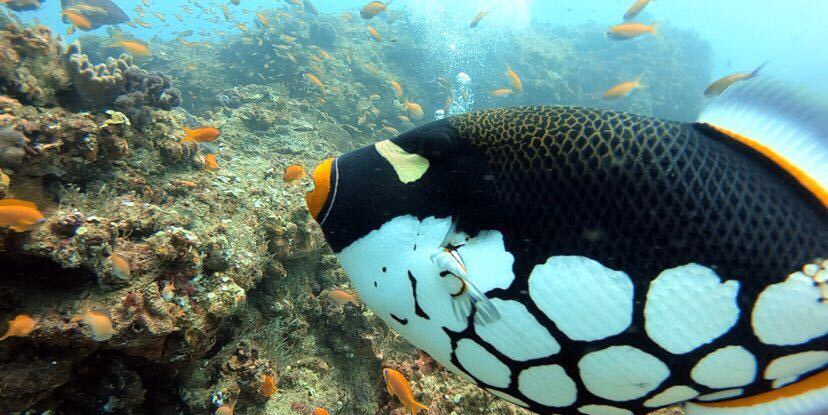}}
\centerline{\scriptsize{(b) MOS:0.8}}
\vspace{4pt}
\end{minipage}
\hfill
\begin{minipage}{0.24\linewidth}
  \centering
\centerline{\includegraphics[width=1\linewidth, height=0.8\linewidth]{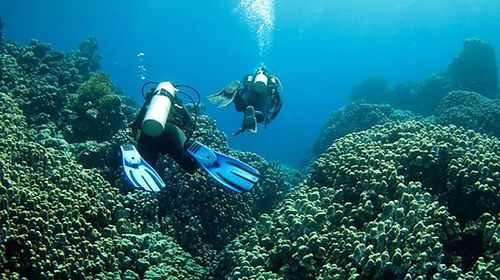}}
\centerline{\scriptsize{(c) MOS:0.7}}
\vspace{4pt}
\end{minipage}
\hfill
\begin{minipage}{0.24\linewidth}
  \centering
\centerline{\includegraphics[width=1\linewidth, height=0.8\linewidth]{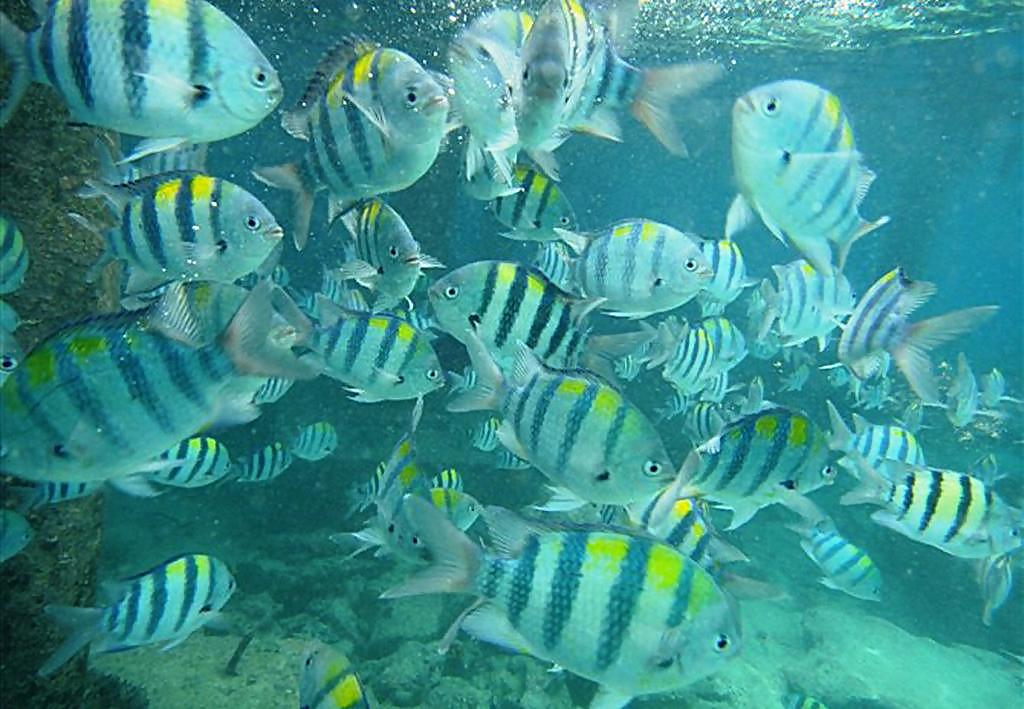}}
\centerline{\scriptsize{(d) MOS:0.6}}
\vspace{4pt}
\end{minipage}

\begin{minipage}{0.24\linewidth}
  \centering
\centerline{\includegraphics[width=1\linewidth, height=0.8\linewidth]{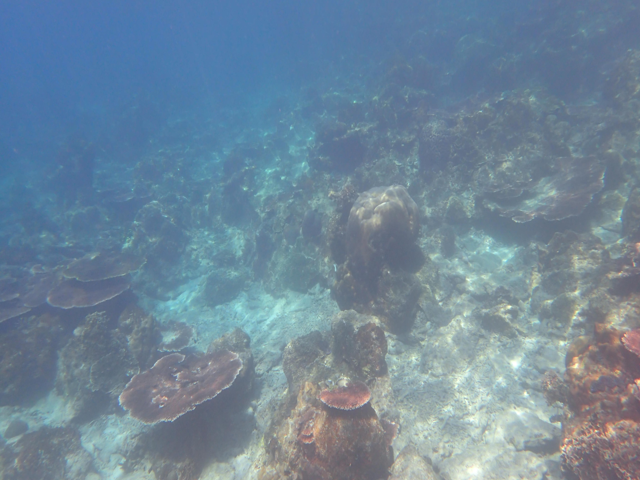}}
\centerline{\scriptsize{(e) MOS:0.5}}
\end{minipage}
\hfill
\begin{minipage}{0.24\linewidth}
  \centering
\centerline{\includegraphics[width=1\linewidth, height=0.8\linewidth]{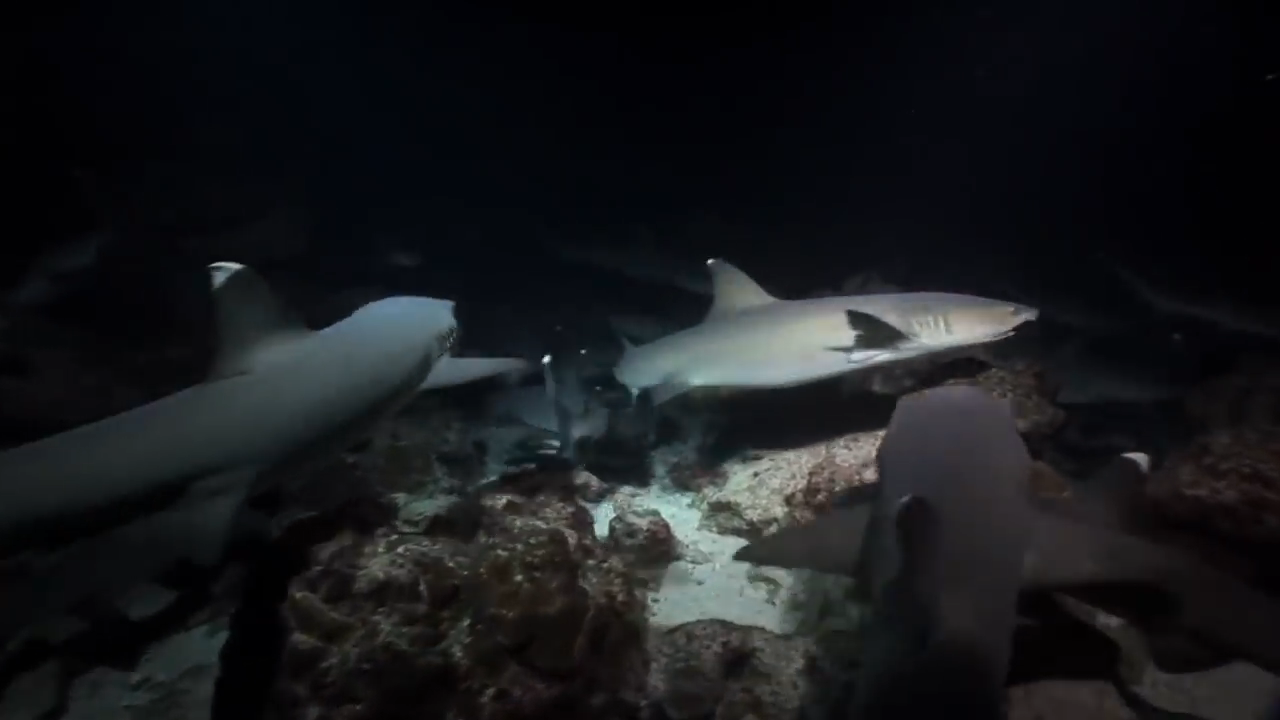}}
\centerline{\scriptsize{(f) MOS:0.4}}
\end{minipage}
\hfill
\begin{minipage}{0.24\linewidth}
  \centering
\centerline{\includegraphics[width=1\linewidth, height=0.8\linewidth]{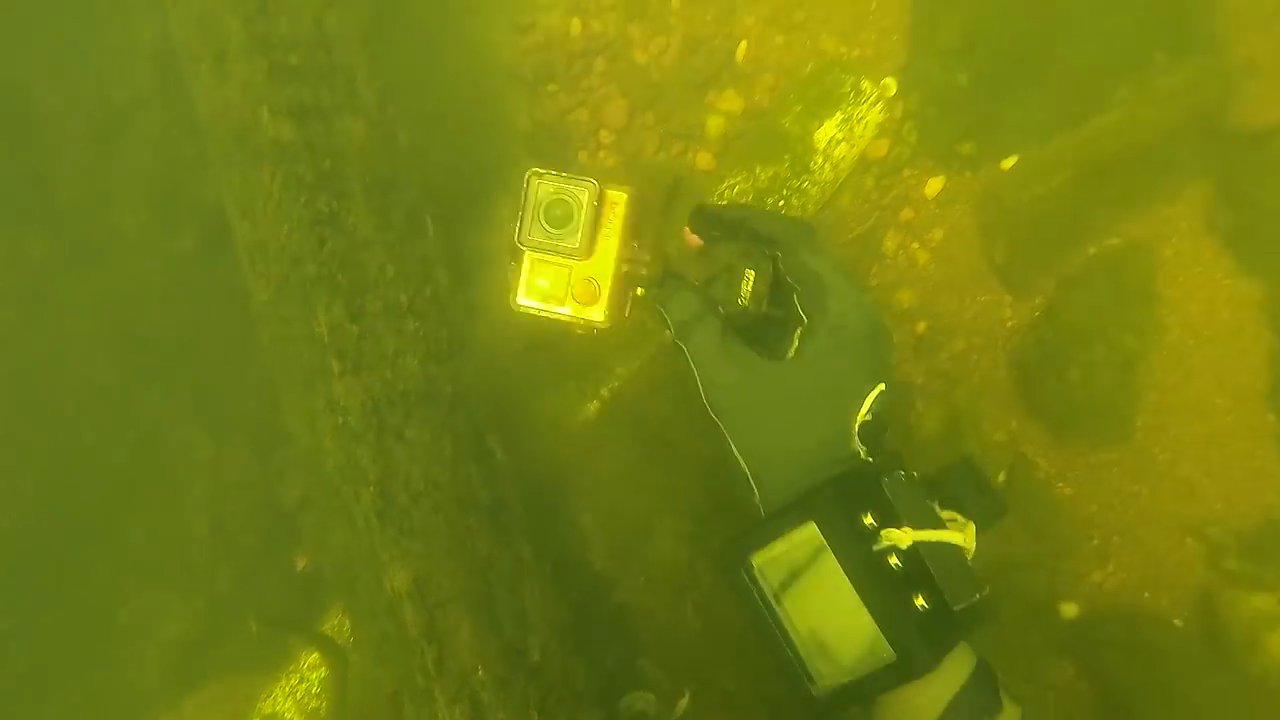}}
\centerline{\scriptsize{(g) MOS:0.3}}
\end{minipage}
\hfill
\begin{minipage}{0.24\linewidth}
  \centering
\centerline{\includegraphics[width=1\linewidth, height=0.8\linewidth]{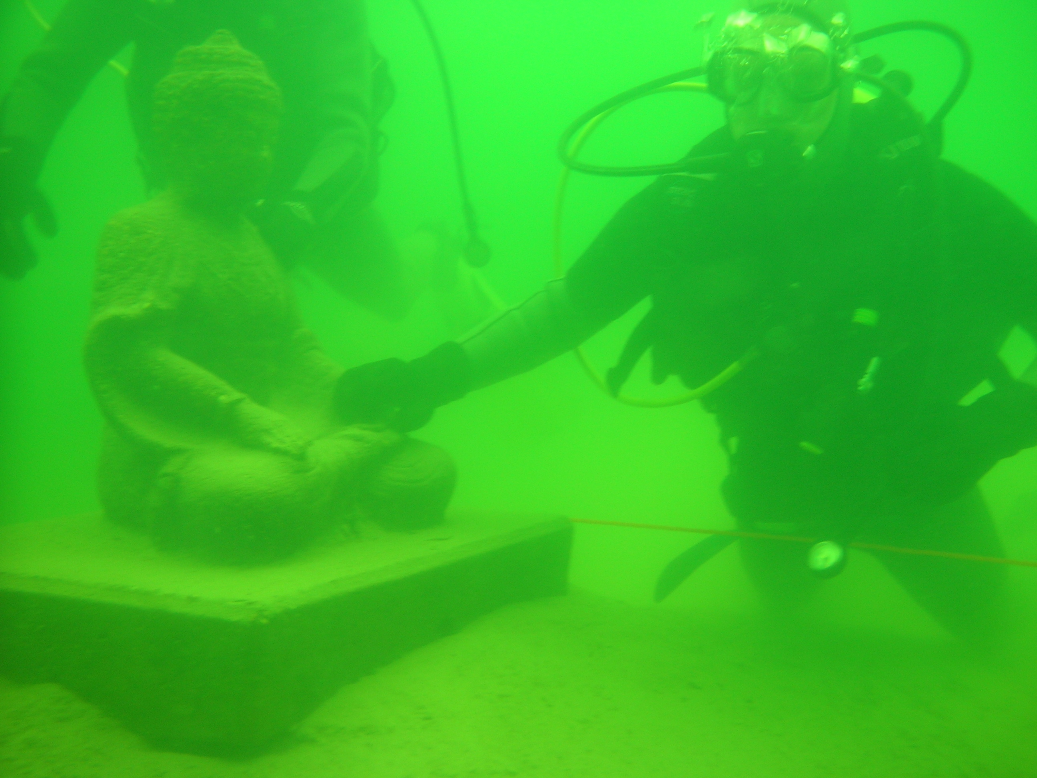}}
\centerline{\scriptsize{(h) MOS:0.2}}
\end{minipage}

\caption{Underwater images of different subjective qualities. A higher MOS value indicates better image quality.}
\label{fig:figs}
\end{figure*}

In addition, we selected eight underwater images of varying subjective quality from the UWIQA database \cite{14}. The evaluated underwater images are displayed in \textbf{Fig. \ref{fig:figs}}, and Table III lists the corresponding objective quality scores predicted by each IQA method. In this table, the symbol ``↑'' following each IQA method indicates that a higher score corresponds to better image quality, whereas the symbol ``↓'' indicates that a lower score signifies better quality. We also ranked the images on the basis of the objective quality scores of each IQA method, with the ranks presented in parentheses.

An analysis of \textbf{Table \ref{tab:my_label}} yields several significant conclusions. First, existing natural image IQA methods fail to provide quality results that are consistent with subjective MOS values; some methods even produce completely contradictory outcomes. For example, the visual quality of the image in Fig. \ref{fig:figs} (h) is the poorest, with an MOS value of 0.2. However, according to the PIQE method, this image ranks first, suggesting that it has the highest quality. Similarly, the NFERM method ranks this image second in quality. The image in Fig. \ref{fig:figs} (g) is deemed the second worst, yet both dipIQ and UNIQUE rate it as the highest-quality image. Additionally, competing underwater IQA methods struggle to accurately differentiate between the quality of the underwater images; for example, UCIQE confuses the quality of Fig. \ref{fig:figs} (b) and (c) as well as Fig. \ref{fig:figs} (d) and (e). UIQM exhibits similar confusion between Fig. \ref{fig:figs} (b) and (c), Fig. \ref{fig:figs} (d) and (e), and Fig. \ref{fig:figs} (g) and (h). Other competing underwater IQA methods also demonstrate a high rate of incorrect quality rankings. In contrast, the proposed UIQI method successfully yields quality rankings that align with subjective MOS values, indicating its strong ability to distinguish underwater image quality.

Overall, the experimental results provide compelling evidence of the superiority of PIGUIQA in evaluating underwater image quality. PIGUIQA not only accurately reflects human subjective perceptions of underwater image quality but also effectively differentiates between various levels of image enhancement. This capability positions PIGUIQA as a powerful tool that can provide reliable guidance for the development and optimization of underwater image enhancement algorithms.

\begin{table*}[t]
\caption{MOS values and objective quality scores predicted by the IQA methods for the underwater images in Fig. \ref{fig:figs}. The number in the brackets refers to the quality rank number according to the MOS or each IQA method}
    \centering
    \resizebox{\textwidth}{!}{
        \begin{tabular}{c| cccccccc}
            \toprule[0.8pt]\bottomrule[0.5pt]
\rowcolor[HTML]{DFDFDF}
            & Fig. \ref{fig:figs} (a)  & Fig. \ref{fig:figs} (b)  & Fig. \ref{fig:figs} (c)  & Fig. \ref{fig:figs} (d)  & Fig. \ref{fig:figs} (e)   & Fig. \ref{fig:figs} (f)  & Fig. \ref{fig:figs} (g)  & Fig. \ref{fig:figs} (h)  \\\hline
MOS (Ground-Truth)    & 0.9 (1)     & 0.8 (2)     & 0.7 (3)     & 0.6 (4)     & 0.5 (5)      & 0.4 (6)     & 0.3 (7)     & 0.2 (8)     \\
\rowcolor[HTML]{DFDFDF} BRISQUE \cite{4} (↑)  & 0.6386 (1)  & 0.6243 (2)  & 0.4974 (8)  & 0.5007 (7)  & 0.5383 (4)   & 0.5124 (5)  & 0.5889 (3)  & 0.5021 (6)  \\
NFERM \cite{5} (↑)    & 0.6883 (3)  & 0.6928 (1)  & 0.5095 (6)  & 0.5286 (4)  & 0.4796 (7)   & 0.5110 (5)  & 0.3685 (8)  & 0.6897 (2)  \\
\rowcolor[HTML]{DFDFDF} NIQE \cite{6} (↓)     & 2.1141 (1)  & 2.6852 (2)  & 6.8129 (4)  & 3.4929 (3)  & 18.5586 (8)  & 7.1692 (6)  & 7.0939 (5)  & 8.3398 (7)  \\
IL-NIQE \cite{7} (↓) & 23.8519 (3) & 17.3602 (1) & 41.3451 (6) & 18.3372 (2) & 58.1182 (8) & 33.2063 (4) & 34.8289 (5) & 55.8821 (7) \\
\rowcolor[HTML]{DFDFDF} SNP-NIQE \cite{8} (↓)  & 3.3420 (1)  & 4.0462 (2)  & 7.4101 (4)  & 6.9708 (3)  & 22.8925 (8)  & 11.7738 (6) & 9.7729 (5)  & 12.5333 (7) \\
PIQE \cite{9} (↓)    & 31.7844 (2) & 34.4559 (4) & 39.1861 (5) & 32.8466 (3) & 46.0535 (6) & 64.5098 (8) & 54.8411 (7) & 7.0233 (1)  \\
\rowcolor[HTML]{DFDFDF} NPQI \cite{10} (↓)     & 3.7873 (1)  & 4.3539 (2)  & 9.1585 (4)  & 8.1506 (3)  & 102.7363 (8) & 15.8425 (6) & 12.0952 (5) & 19.8938 (7) \\
dipIQ \cite{11} (↑)   & -7.0445 (5) & -3.7943 (4) & -2.1332 (2) & -9.3835 (7) & -2.3022 (3) & -7.3323 (6) & -1.8703 (1) & -9.8172 (8) \\
\rowcolor[HTML]{DFDFDF} HyperIQA \cite{12} (↑) & 0.6487 (2)  & 0.6605 (1)  & 0.5209 (5)  & 0.5551 (3)  & 0.3939 (7)   & 0.4559 (6)  & 0.5309 (4)  & 0.3667 (8)  \\
UNIQUE \cite{13} (↑)   & 0.8216 (2)  & 0.8000 (4)  & 0.6714 (7)  & 0.6197 (8)  & 0.7219 (5)   & 0.8141 (3)  & 0.9512 (1)  & 0.6983 (6)  \\\hline
\rowcolor[HTML]{DFDFDF} UCIQE \cite{15} (↑)    & 35.8752 (1) & 32.2376 (3) & 32.7884 (2) & 30.6955 (5) & 31.9299 (4) & 27.0893 (6) & 26.3908 (7) & 19.0256 (8) \\
UIQM \cite{16} (↑)      & 1.7310 (1)  & 1.5228 (3)  & 1.6125 (2)  & 1.2535 (5)  & 1.4972 (4)   & 1.1841 (6)  & 0.5913 (8)  & 0.6035 (7)  \\
\rowcolor[HTML]{DFDFDF} CCF \cite{17} (↑)      & 26.9893 (4) & 34.8443 (2) & 37.9844 (1) & 23.7656 (5) & 31.2540 (3)  & 14.1339 (7) & 21.6344 (6) & 8.1107 (8)  \\
FDUM \cite{14} (↑)     & 1.0136 (2)  & 0.7822 (3)  & 1.1714 (1)  & 0.4152 (5)  & 0.7635 (4)   & 0.1921 (7)  & 0.2699 (6)  & 0.1176 (8)  \\
\rowcolor[HTML]{DFDFDF} PIGUIQA (ours)  (↑)       & 0.7862 (1)  & 0.6830 (2)  & 0.6055 (3)  & 0.5600 (4)  & 0.5272 (5)   & 0.4554 (6)  & 0.3680 (7)  & 0.2581 (8)\\
\toprule[0.5pt]\bottomrule[0.8pt]
        \end{tabular}
    }
    \label{tab:my_label}
\end{table*}

\subsection{Visualization of Local Distortion-Aware Functions}

\begin{figure*}[t]

\begin{minipage}{0.195\linewidth}
  \centering
\centerline{\includegraphics[width=1\linewidth, height=0.8\linewidth]{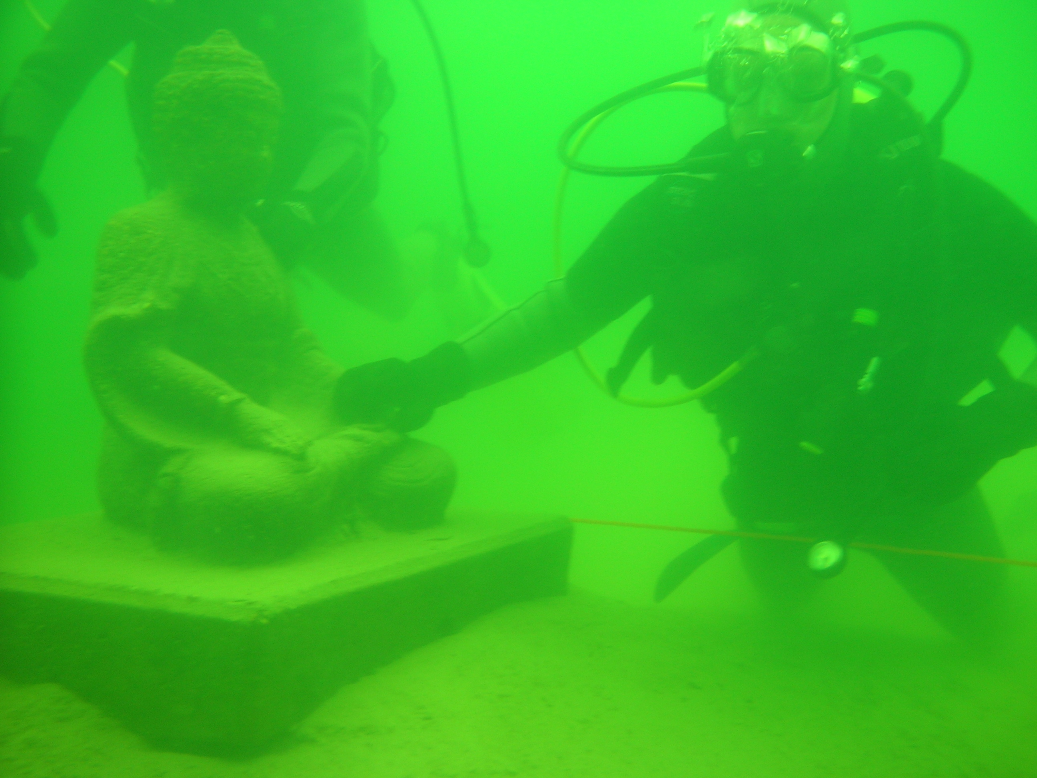}}
\vspace{4pt}
\end{minipage}
\hfill
\begin{minipage}{0.195\linewidth}
  \centering
\centerline{\includegraphics[width=1\linewidth, height=0.8\linewidth]{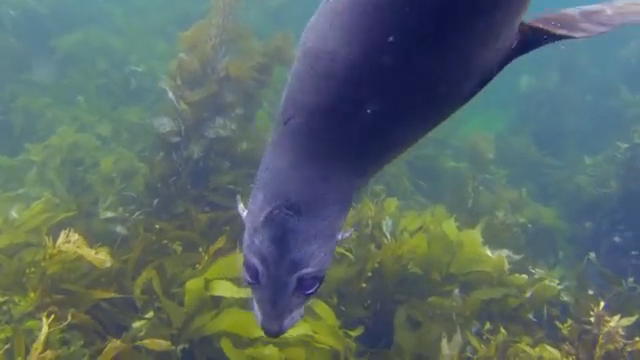}}
\vspace{4pt}
\end{minipage}
\hfill
\begin{minipage}{0.195\linewidth}
  \centering
\centerline{\includegraphics[width=1\linewidth, height=0.8\linewidth]{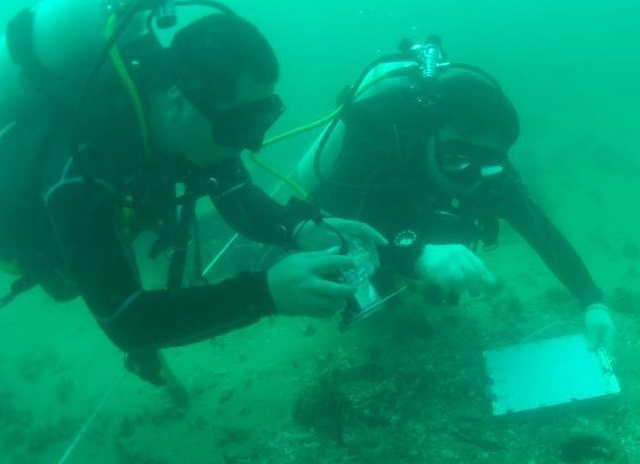}}
\vspace{4pt}
\end{minipage}
\hfill
\begin{minipage}{0.195\linewidth}
  \centering
\centerline{\includegraphics[width=1\linewidth, height=0.8\linewidth]{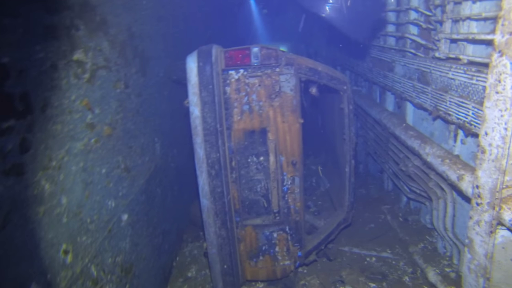}}
\vspace{4pt}
\end{minipage}
\hfill
\begin{minipage}{0.195\linewidth}
  \centering
\centerline{\includegraphics[width=1\linewidth, height=0.8\linewidth]{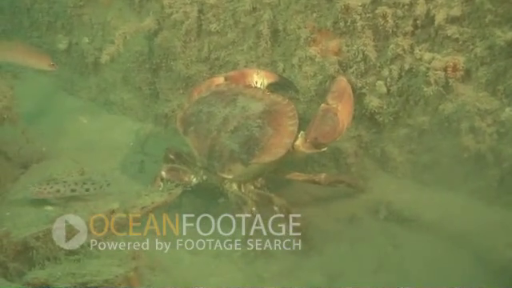}}
\vspace{4pt}
\end{minipage}

\begin{minipage}{0.195\linewidth}
  \centering
\centerline{\includegraphics[width=1\linewidth, height=0.8\linewidth]{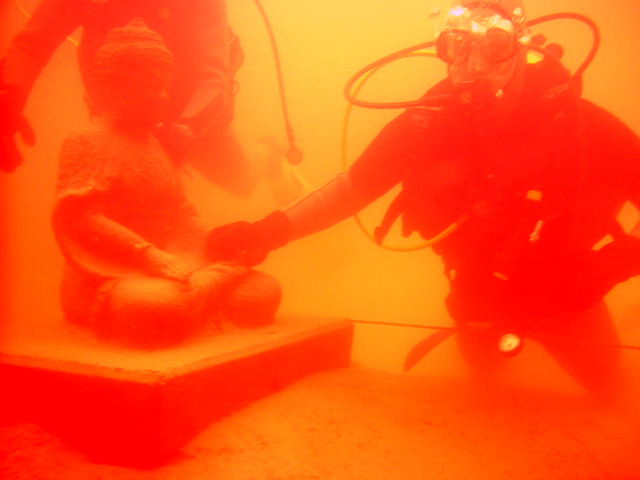}}
\vspace{4pt}
\end{minipage}
\hfill
\begin{minipage}{0.195\linewidth}
  \centering
\centerline{\includegraphics[width=1\linewidth, height=0.8\linewidth]{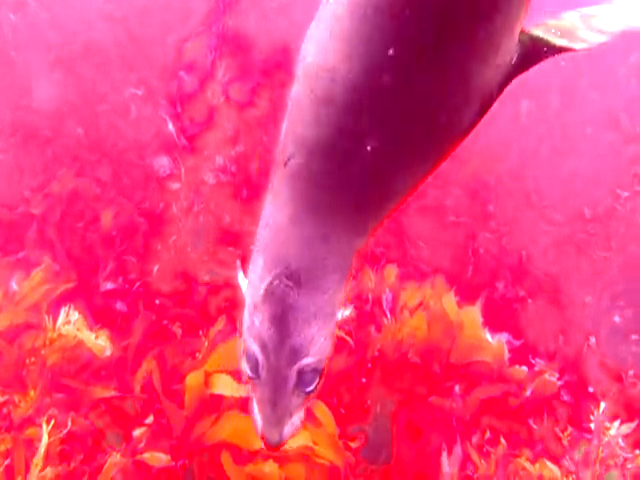}}
\vspace{4pt}
\end{minipage}
\hfill
\begin{minipage}{0.195\linewidth}
  \centering
\centerline{\includegraphics[width=1\linewidth, height=0.8\linewidth]{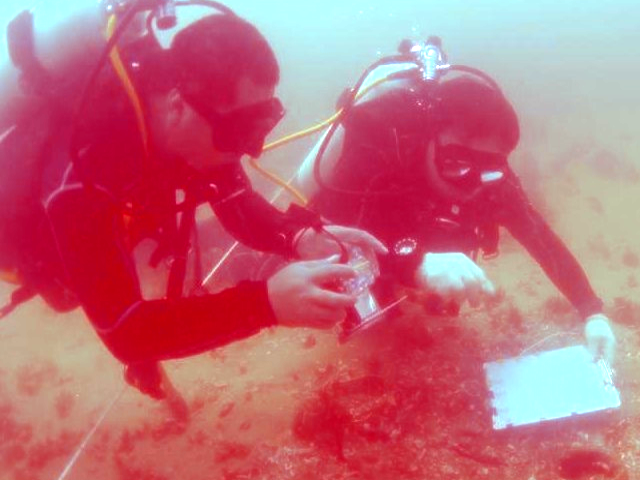}}
\vspace{4pt}
\end{minipage}
\hfill
\begin{minipage}{0.195\linewidth}
  \centering
\centerline{\includegraphics[width=1\linewidth, height=0.8\linewidth]{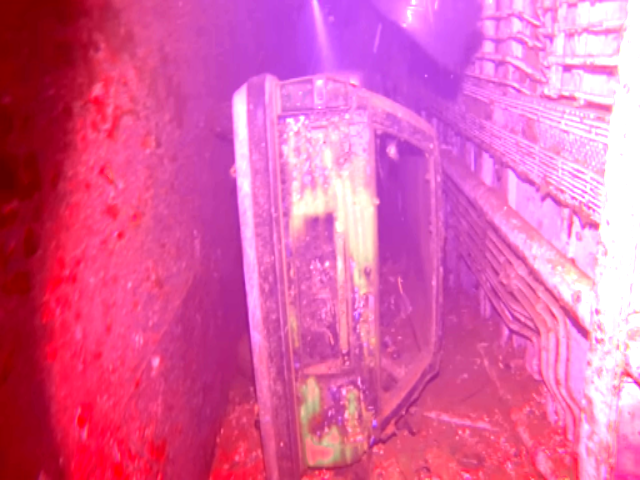}}
\vspace{4pt}
\end{minipage}
\hfill
\begin{minipage}{0.195\linewidth}
  \centering
\centerline{\includegraphics[width=1\linewidth, height=0.8\linewidth]{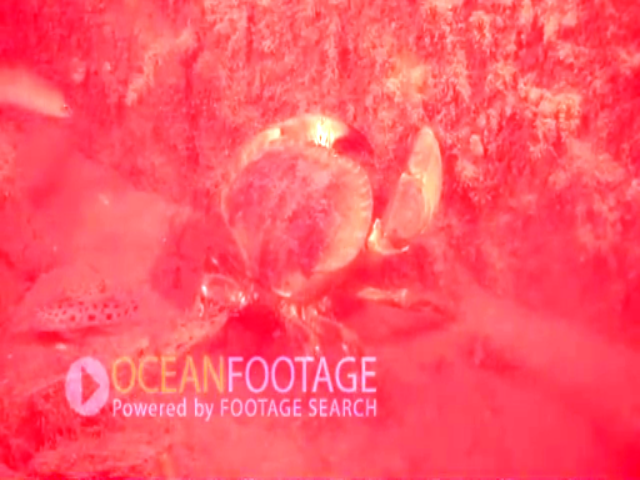}}
\vspace{4pt}
\end{minipage}

\begin{minipage}{0.195\linewidth}
  \centering
\centerline{\includegraphics[width=1\linewidth, height=0.8\linewidth]{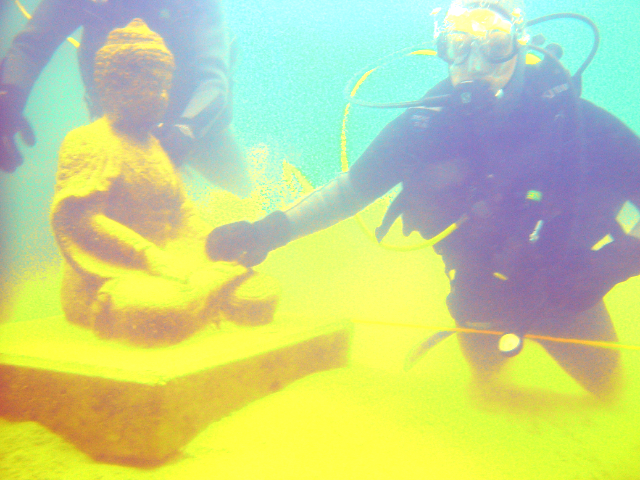}}
\end{minipage}
\hfill
\begin{minipage}{0.195\linewidth}
  \centering
\centerline{\includegraphics[width=1\linewidth, height=0.8\linewidth]{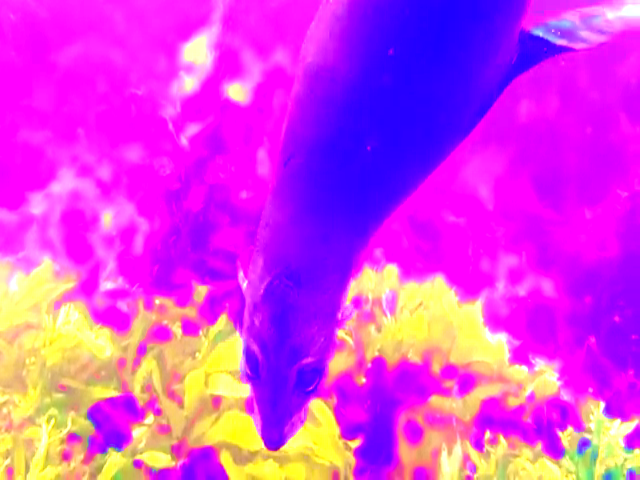}}
\end{minipage}
\hfill
\begin{minipage}{0.195\linewidth}
  \centering
\centerline{\includegraphics[width=1\linewidth, height=0.8\linewidth]{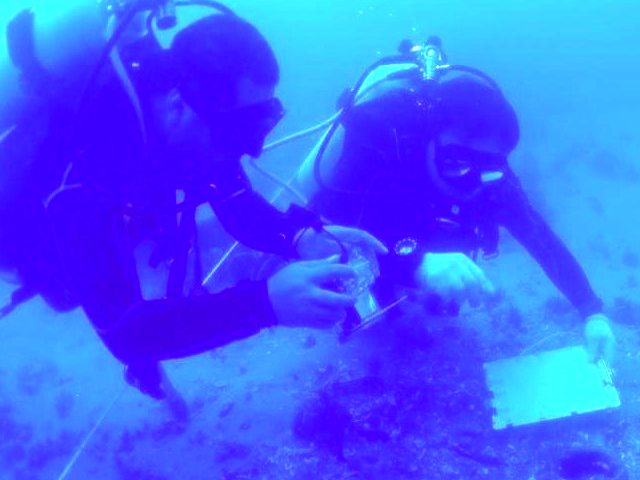}}
\end{minipage}
\hfill
\begin{minipage}{0.195\linewidth}
  \centering
\centerline{\includegraphics[width=1\linewidth, height=0.8\linewidth]{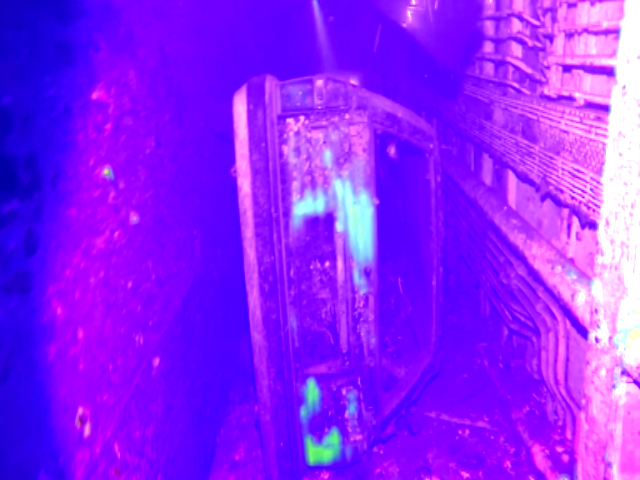}}
\end{minipage}
\hfill
\begin{minipage}{0.195\linewidth}
  \centering
\centerline{\includegraphics[width=1\linewidth, height=0.8\linewidth]{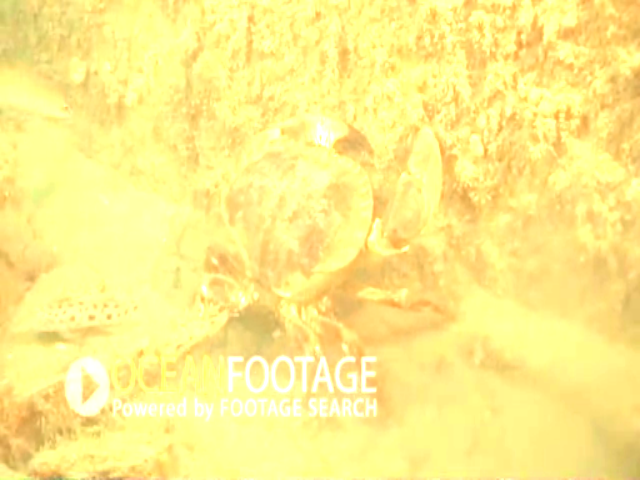}}
\end{minipage}

\caption{Visualization of local distortion-aware functions. Row 1: original underwater images $I$. Row 2: transmission attenuation distortion-aware map ($f_1(I)$). Row 3: Backwards scattering distortion-aware map ($f_2(I)$).}
\label{fig:aware}

\end{figure*}

To illustrate the perceptual capabilities of the local transmission attenuation distortion-aware network ($f_1$) and the local backwards scattering distortion-aware network ($f_2$) in relation to different distortion types, we conducted a visualization of their network outputs. As shown in \textbf{Fig. \ref{fig:aware}}, the output image from $f_1$ exhibits a predominantly red hue, whereas the output from $f_2$ leans more towards green. This observation indicates that both networks effectively perceive transmission attenuation distortion and scattering distortion, demonstrating precise sensitivity to their respective distortion types.

From a physical perspective, this phenomenon is closely tied to the characteristics of light propagation underwater. In aquatic environments, light of different wavelengths travels varying distances; specifically, red light, with its longer wavelength, is absorbed more readily by water, leading to quicker attenuation during transmission. In contrast, blue and green light, with shorter wavelengths, can penetrate greater distances, resulting in enhanced visibility in the background. Consequently, transmission attenuation distortion primarily manifests in the attenuation of red light, necessitating special attention to the red regions of the output image. Conversely, scattering distortion arises mainly from the scattering of background light, which is predominantly composed of blue and green wavelengths, on which the network must focus in the output.

Specifically, $f_1$ enhances its sensitivity to red light, thereby improving its ability to capture signal distortions caused by attenuation during transmission. This characteristic endows $f_1$ with heightened sensitivity in scenarios where red light experiences significant attenuation. In contrast, $f_2$ demonstrates increased sensitivity to blue and green light, effectively detecting distortion noise generated by scattering from background light, thus enabling a distinction between valuable information and extraneous scattered signals.

The results from this visualization not only validate the targeted design and effectiveness of $f_1$ and $f_2$ but also confirm the scientific principles governing underwater light propagation from a perceptual mechanism perspective. Through comparative analysis of the red and blue–green regions in the network outputs, we further establish that $f_1$ and $f_2$ are capable of providing effective perceptual abilities tailored to different sources of distortion.

Thus, the incorporation of underwater imaging prior knowledge, particularly regarding the perception of transmission attenuation and backwards scattering distortions, significantly enhances the model's capacity to assess underwater image quality. This finding further substantiates the effectiveness of our approach: by integrating the physical model of underwater imaging, the model can more accurately identify and predict quality issues in underwater images, thereby improving the precision and reliability of image quality assessment.

\section{Conclusion}
\label{sec:conclu}

In this paper, we propose a novel physical imaging guided framework for UIQA, PIGUIQA. By modeling underwater image degradation as a combined effect of direct transmission attenuation and backward scattering, we establish physics-based distortion metrics that quantitatively characterize their perceptual impacts. Moreover, local perceptual modules utilizing a neighborhood attention mechanism are designed to capture subtle spatial features and prioritize perceptually sensitive areas. The framework also introduces a global perceptual aggregator that holistically fuses scene semantics with distortion characteristics to ensure accurate quality score prediction. Extensive experiments across diverse underwater benchmarks validate that PIGUIQA achieves state-of-the-art performance, demonstrating superior correlation with human subjective judgments and exceptional cross-dataset generalization capabilities.

\small
\bibliographystyle{ieeenat_fullname}
\bibliography{main}

@INPROCEEDINGS{1,
  author={Wen, Junjie and Cui, Jinqiang and Zhao, Zhenjun and Yan, Ruixin and Gao, Zhi and Dou, Lihua and Chen, Ben M.},
  booktitle={2023 IEEE International Conference on Robotics and Automation (ICRA)},
  title={SyreaNet: A Physically Guided Underwater Image Enhancement Framework Integrating Synthetic and Real Images},
  year={2023},
  volume={},
  number={},
  pages={5177-5183}
}

@INPROCEEDINGS{2,
  author={Akkaynak, Derya and Treibitz, Tali},
  booktitle={2018 IEEE/CVF Conference on Computer Vision and Pattern Recognition (CVPR)},
  title={A Revised Underwater Image Formation Model},
  year={2018},
  volume={},
  number={},
  pages={6723-6732}
}

@ARTICLE{3,
  author={Liu, Yutao and Gu, Ke and Cao, Jingchao and Wang, Shiqi and Zhai, Guangtao and Dong, Junyu and Kwong, Sam},
  journal={IEEE Transactions on Multimedia},
  title={UIQI: A Comprehensive Quality Evaluation Index for Underwater Images},
  year={2024},
  volume={26},
  number={},
  pages={2560-2573}
}

@ARTICLE{4,
  author={Mittal, Anish and Moorthy, Anush Krishna and Bovik, Alan Conrad},
  journal={IEEE Transactions on Image Processing},
  title={No-Reference Image Quality Assessment in the Spatial Domain},
  year={2012},
  volume={21},
  number={12},
  pages={4695-4708}
}

@ARTICLE{5,
  author={Gu, Ke and Zhai, Guangtao and Yang, Xiaokang and Zhang, Wenjun},
  journal={IEEE Transactions on Multimedia},
  title={Using Free Energy Principle For Blind Image Quality Assessment},
  year={2015},
  volume={17},
  number={1},
  pages={50-63}
}

@ARTICLE{6,
  author={Mittal, Anish and Soundararajan, Rajiv and Bovik, Alan C.},
  journal={IEEE Signal Processing Letters},
  title={Making a “Completely Blind” Image Quality Analyzer},
  year={2013},
  volume={20},
  number={3},
  pages={209-212}
}

@ARTICLE{7,
  author={Zhang, Lin and Zhang, Lei and Bovik, Alan C.},
  journal={IEEE Transactions on Image Processing},
  title={A Feature-Enriched Completely Blind Image Quality Evaluator},
  year={2015},
  volume={24},
  number={8},
  pages={2579-2591}
}

@ARTICLE{8,
  author={Liu, Yutao and Gu, Ke and Zhang, Yongbing and Li, Xiu and Zhai, Guangtao and Zhao, Debin and Gao, Wen},
  journal={IEEE Transactions on Circuits and Systems for Video Technology},
  title={Unsupervised Blind Image Quality Evaluation via Statistical Measurements of Structure, Naturalness, and Perception},
  year={2020},
  volume={30},
  number={4},
  pages={929-943}
}

@INPROCEEDINGS{9,
  author={Venkatanath N and Praneeth D and Maruthi Chandrasekhar Bh and Channappayya, Sumohana S. and Medasani, Swarup S.},
  booktitle={2015 Twenty First National Conference on Communications (NCC)},
  title={Blind image quality evaluation using perception based features},
  year={2015},
  volume={},
  number={},
  pages={1-6}
}

@ARTICLE{10,
  author={Liu, Yutao and Gu, Ke and Li, Xiu and Zhang, Yongbing},
  title={Blind Image Quality Assessment by Natural Scene Statistics and Perceptual Characteristics},
  year={2020},
  volume={16},
  number={3},
  journal={ACM Transactions on Multimedia Computing, Communications, and Applications},
  pages={1-91}
}

@ARTICLE{11,
  author={Ma, Kede and Liu, Wentao and Liu, Tongliang and Wang, Zhou and Tao, Dacheng},
  journal={IEEE Transactions on Image Processing},
  title={dipIQ: Blind Image Quality Assessment by Learning-to-Rank Discriminable Image Pairs},
  year={2017},
  volume={26},
  number={8},
  pages={3951-3964}
}

@INPROCEEDINGS{12,
  author={Su, Shaolin and Yan, Qingsen and Zhu, Yu and Zhang, Cheng and Ge, Xin and Sun, Jinqiu and Zhang, Yanning},
  booktitle={2020 IEEE/CVF Conference on Computer Vision and Pattern Recognition (CVPR)},
  title={Blindly Assess Image Quality in the Wild Guided by a Self-Adaptive Hyper Network},
  year={2020},
  volume={},
  number={},
  pages={3664-3673}
}

@ARTICLE{13,
  author={Zhang, Weixia and Ma, Kede and Zhai, Guangtao and Yang, Xiaokang},
  journal={IEEE Transactions on Image Processing},
  title={Uncertainty-Aware Blind Image Quality Assessment in the Laboratory and Wild},
  year={2021},
  volume={30},
  number={},
  pages={3474-3486}
}

@ARTICLE{14,
  title={A reference-free underwater image quality assessment metric in frequency domain},
  journal={Signal Processing: Image Communication},
  volume={94},
  pages={116218},
  year={2021},
  author={Ning Yang and Qihang Zhong and Kun Li and Runmin Cong and Yao Zhao and Sam Kwong}
}

@ARTICLE{15,
  author={Yang, Miao and Sowmya, Arcot},
  journal={IEEE Transactions on Image Processing},
  title={An Underwater Color Image Quality Evaluation Metric},
  year={2015},
  volume={24},
  number={12},
  pages={6062-6071}
}

@ARTICLE{16,
  author={Panetta, Karen and Gao, Chen and Agaian, Sos},
  journal={IEEE Journal of Oceanic Engineering},
  title={Human-Visual-System-Inspired Underwater Image Quality Measures},
  year={2016},
  volume={41},
  number={3},
  pages={541-551}
}

@ARTICLE{17,
  title={An imaging-inspired no-reference underwater color image quality assessment metric},
  journal={Computers \& Electrical Engineering},
  volume={70},
  pages={904-913},
  year={2018},
  author={Yan Wang and Na Li and Zongying Li and Zhaorui Gu and Haiyong Zheng and Bing Zheng and Mengnan Sun},
}

@INPROCEEDINGS{18,
  author={Huang, Shirui and Wang, Keyan and Liu, Huan and Chen, Jun and Li, Yunsong},
  booktitle={2023 IEEE/CVF Conference on Computer Vision and Pattern Recognition (CVPR)},
  title={Contrastive Semi-Supervised Learning for Underwater Image Restoration via Reliable Bank},
  year={2023},
  volume={},
  number={},
  pages={18145-18155}
}

@INPROCEEDINGS{19,
  author={Zhao, Chen and Cai, Weiling and Dong, Chenyu and Hu, Chengwei},
  booktitle={2024 IEEE/CVF Conference on Computer Vision and Pattern Recognition (CVPR)},
  title={Wavelet-based Fourier Information Interaction with Frequency Diffusion Adjustment for Underwater Image Restoration},
  year={2024},
  volume={},
  number={},
  pages={8281-8291}
}

@ARTICLE{20,
  author={Zhou Wang and Bovik, A.C. and Sheikh, H.R. and Simoncelli, E.P.},
  journal={IEEE Transactions on Image Processing},
  title={Image quality assessment: from error visibility to structural similarity},
  year={2004},
  volume={13},
  number={4},
  pages={600-612}
}

@ARTICLE{21,
  author={Zhang, Lin and Zhang, Lei and Mou, Xuanqin and Zhang, David},
  journal={IEEE Transactions on Image Processing},
  title={FSIM: A Feature Similarity Index for Image Quality Assessment},
  year={2011},
  volume={20},
  number={8},
  pages={2378-2386}
}

@ARTICLE{22,
  author={Hou, Guojia and Li, Yuxuan and Yang, Huan and Li, Kunqian and Pan, Zhenkuan},
  title={UID2021: An Underwater Image Dataset for Evaluation of No-Reference Quality Assessment Metrics},
  year={2023},
  volume={19},
  number={4},
  pages={1-24},
  journal={ACM Transactions on Multimedia Computing, Communications, and Applications}
}

@ARTICLE{23,
  author={Saad, Michele A. and Bovik, Alan C. and Charrier, Christophe},
  journal={IEEE Transactions on Image Processing},
  title={Blind Image Quality Assessment: A Natural Scene Statistics Approach in the DCT Domain},
  year={2012},
  volume={21},
  number={8},
  pages={3339-3352}
}

@ARTICLE{24,
  author={Li, Qiaohong and Lin, Weisi and Xu, Jingtao and Fang, Yuming},
  journal={IEEE Transactions on Multimedia},
  title={Blind Image Quality Assessment Using Statistical Structural and Luminance Features},
  year={2016},
  volume={18},
  number={12},
  pages={2457-2469}
}

@ARTICLE{25,
  author={Li, Leida and Xia, Wenhan and Lin, Weisi and Fang, Yuming and Wang, Shiqi},
  journal={IEEE Transactions on Multimedia},
  title={No-Reference and Robust Image Sharpness Evaluation Based on Multiscale Spatial and Spectral Features},
  year={2017},
  volume={19},
  number={5},
  pages={1030-1040}
}

@ARTICLE{26,
  author={Moorthy, Anush Krishna and Bovik, Alan Conrad},
  journal={IEEE Transactions on Image Processing},
  title={Blind Image Quality Assessment: From Natural Scene Statistics to Perceptual Quality},
  year={2011},
  volume={20},
  number={12},
  pages={3350-3364}
}

@ARTICLE{27,
  author={Zhang, Weixia and Ma, Kede and Yan, Jia and Deng, Dexiang and Wang, Zhou},
  journal={IEEE Transactions on Circuits and Systems for Video Technology},
  title={Blind Image Quality Assessment Using a Deep Bilinear Convolutional Neural Network},
  year={2020},
  volume={30},
  number={1},
  pages={36-47}
}

@ARTICLE{28,
  author={Wu, Qingbo and Wang, Lei and Ngan, King Ngi and Li, Hongliang and Meng, Fanman and Xu, Linfeng},
  journal={IEEE Transactions on Circuits and Systems for Video Technology},
  title={Subjective and Objective De-Raining Quality Assessment Towards Authentic Rain Image},
  year={2020},
  volume={30},
  number={11},
  pages={3883-3897}
}

@ARTICLE{29,
  author={Jiang, Qiuping and Gu, Yuese and Li, Chongyi and Cong, Runmin and Shao, Feng},
  journal={IEEE Transactions on Circuits and Systems for Video Technology},
  title={Underwater Image Enhancement Quality Evaluation: Benchmark Dataset and Objective Metric},
  year={2022},
  volume={32},
  number={9},
  pages={5959-5974}
}

@ARTICLE{30,
  author={Jiang, Qiuping and Yi, Xiao and Ouyang, Li and Zhou, Jingchun and Wang, Zhihua},
  journal={IEEE Transactions on Circuits and Systems for Video Technology},
  title={Towards Dimension-Enriched Underwater Image Quality Assessment},
  year={2024},
  volume={},
  number={},
  pages={1-1}
}

@ARTICLE{31,
  author={Bosse, Sebastian and Maniry, Dominique and Müller, Klaus-Robert and Wiegand, Thomas and Samek, Wojciech},
  journal={IEEE Transactions on Image Processing},
  title={Deep Neural Networks for No-Reference and Full-Reference Image Quality Assessment},
  year={2018},
  volume={27},
  number={1},
  pages={206-219}
}

@ARTICLE{32,
  author={Gu, Ke and Tao, Dacheng and Qiao, Jun-Fei and Lin, Weisi},
  journal={IEEE Transactions on Neural Networks and Learning Systems},
  title={Learning a No-Reference Quality Assessment Model of Enhanced Images With Big Data},
  year={2018},
  volume={29},
  number={4},
  pages={1301-1313}
}

@ARTICLE{33,
  author={Ghadiyaram, Deepti and Bovik, Alan C.},
  title={Perceptual quality prediction on authentically distorted images using a bag of features approach},
  journal={Journal of Vision},
  volume={17},
  number={1},
  pages={32},
  year={2017}
}

@ARTICLE{34,
  author={Yang, Miao and Sowmya, Arcot},
  journal={IEEE Transactions on Image Processing},
  title={An Underwater Color Image Quality Evaluation Metric},
  year={2015},
  volume={24},
  number={12},
  pages={6062-6071}
}

@ARTICLE{35,
  author={Panetta, Karen and Gao, Chen and Agaian, Sos},
  journal={IEEE Journal of Oceanic Engineering},
  title={Human-Visual-System-Inspired Underwater Image Quality Measures},
  year={2016},
  volume={41},
  number={3},
  pages={541-551}
}

@INPROCEEDINGS{36,
  author={Akkaynak, Derya and Treibitz, Tali},
  booktitle={2019 IEEE/CVF Conference on Computer Vision and Pattern Recognition (CVPR)},
  title={Sea-Thru: A Method for Removing Water From Underwater Images},
  year={2019},
  volume={},
  number={},
  pages={1682-1691}
}

@INPROCEEDINGS{37,
  author={He, Kaiming and Zhang, Xiangyu and Ren, Shaoqing and Sun, Jian},
  booktitle={2016 IEEE Conference on Computer Vision and Pattern Recognition (CVPR)},
  title={Deep Residual Learning for Image Recognition},
  year={2016},
  volume={},
  number={},
  pages={770-778}
}

@ARTICLE{38,
  author={Talebi, Hossein and Milanfar, Peyman},
  journal={IEEE Transactions on Image Processing},
  title={NIMA: Neural Image Assessment},
  year={2018},
  volume={27},
  number={8},
  pages={3998-4011}
}

@ARTICLE{39,
  author={He, Kaiming and Sun, Jian and Tang, Xiaoou},
  journal={IEEE Transactions on Pattern Analysis and Machine Intelligence},
  title={Single Image Haze Removal Using Dark Channel Prior},
  year={2011},
  volume={33},
  number={12},
  pages={2341-2353}
}

@ARTICLE{40,
  author={Zheng, Yannan and Chen, Weiling and Lin, Rongfu and Zhao, Tiesong and Le Callet, Patrick},
  journal={IEEE Transactions on Image Processing},
  title={UIF: An Objective Quality Assessment for Underwater Image Enhancement},
  year={2022},
  volume={31},
  number={},
  pages={5456-5468}
}

@ARTICLE{41,
  author={Guo, Pengfei and Liu, Hantao and Zeng, Delu and Xiang, Tao and Li, Leida and Gu, Ke},
  journal={IEEE Transactions on Multimedia},
  title={An Underwater Image Quality Assessment Metric},
  year={2023},
  volume={25},
  number={},
  pages={5093-5106}
}

@article{42,
  title={Twice Mixing: A rank learning based quality assessment approach for underwater image enhancement},
  journal={Signal Processing: Image Communication},
  volume={102},
  pages={116622},
  year={2022},
  author={Zhenqi Fu and Xueyang Fu and Yue Huang and Xinghao Ding}
}

@ARTICLE{43,
  author={Wang, Zheyin and Shen, Liquan and Wang, Zhengyong and Lin, Yufei and Jin, Yanliang},
  journal={IEEE Transactions on Circuits and Systems for Video Technology},
  title={Generation-Based Joint Luminance-Chrominance Learning for Underwater Image Quality Assessment},
  year={2023},
  volume={33},
  number={3},
  pages={1123-1139}
}

@ARTICLE{44,
  author={Guo, Pengfei and He, Lang and Liu, Shuangyin and Zeng, Delu and Liu, Hantao},
  journal={IEEE Transactions on Multimedia},
  title={Underwater Image Quality Assessment: Subjective and Objective Methods},
  year={2022},
  volume={24},
  number={},
  pages={1980-1989}
}

@INPROCEEDINGS{45,
  author={Varghese, Nisha and Kumar, Ashish and Rajagopalan, A. N.},
  booktitle={2023 IEEE/CVF International Conference on Computer Vision (ICCV)},
  title={Self-supervised Monocular Underwater Depth Recovery, Image Restoration, and a Real-sea Video Dataset},
  year={2023},
  volume={},
  number={},
  pages={12214-12224}
}

@INPROCEEDINGS{46,
  author={Lian, Shijie and Li, Hua and Cong, Runmin and Li, Suqi and Zhang, Wei and Kwong, Sam},
  booktitle={2023 IEEE/CVF International Conference on Computer Vision (ICCV)},
  title={WaterMask: Instance Segmentation for Underwater Imagery},
  year={2023},
  volume={},
  number={},
  pages={1305-1315}
}

@INPROCEEDINGS{47,
  author={Roy, Subhadeep and Mitra, Shankhanil and Biswas, Soma and Soundararajan, Rajiv},
  booktitle={2023 IEEE/CVF International Conference on Computer Vision (ICCV)},
  title={Test Time Adaptation for Blind Image Quality Assessment},
  year={2023},
  volume={},
  number={},
  pages={16696-16705}
}

@INPROCEEDINGS{48,
  author={Xue, Wufeng and Mou, Xuanqin and Zhang, Lei and Feng, Xiangchu},
  booktitle={2013 IEEE/CVF International Conference on Computer Vision (ICCV)},
  title={Perceptual Fidelity Aware Mean Squared Error},
  year={2013},
  volume={},
  number={},
  pages={705-712}
}

@INPROCEEDINGS{49,
  author={Zheng, Heliang and Yang, Huan and Fu, Jianlong and Zha, Zheng-Jun and Luo, Jiebo},
  booktitle={2021 IEEE/CVF International Conference on Computer Vision (ICCV)},
  title={Learning Conditional Knowledge Distillation for Degraded-Reference Image Quality Assessment},
  year={2021},
  volume={},
  number={},
  pages={10222-10231}
}

@INPROCEEDINGS{50,
  title={Data-Efficient Image Quality Assessment with Attention-Panel Decoder},
  author={Guanyi Qin and Runze Hu and Yutao Liu and Xiawu Zheng and Haotian Liu and Xiu Li and Yan Zhang},
  booktitle={Proceedings of the AAAI Conference on Artificial Intelligence (AAAI)},
  year={2023},
  pages={2091-2100}
}

\end{document}